\documentclass[oneside]{jfm}
\usepackage[dvips]{graphicx}
\usepackage{natbib}
\usepackage{upmath}
\usepackage{bm}
\usepackage{multirow}
\usepackage{amssymb, amsmath}
\renewcommand\Re{\mbox{Re}}

\def\mathi{\mathrm{i}}

\title{New wave generation}
\author[Mercier, Martinand, Mathur, Gostiaux, Peacock and Dauxois]
{
M\ls A\ls T\ls T\ls H\ls I\ls E\ls U\ns J.\ns M\ls E\ls R\ls C\ls I\ls E\ls R$^{(1)}$\ns,   D\ls E\ls N\ls I\ls S\ns M\ls A\ls R\ls T\ls I\ls N\ls A\ls N\ls D$^{(2)}$\ns, M\ls A\ls N\ls I\ls K\ls A\ls N\ls
D\ls A\ls N\ns M\ls A\ls T\ls H\ls U\ls R$^{(3)}$\ns, L\ls O\ls U\ls I\ls S\ns G\ls O\ls S\ls T\ls I\ls A\ls U\ls X$^{(4)}$\ns, T\ls H\ls O\ls M\ls A\ls S\ns P\ls E\ls A\ls C\ls O\ls C\ls K$^{(3)}$
\ns \and T\ls H\ls I\ls E\ls R\ls R\ls Y\ns  D\ls A\ls U\ls X\ls O\ls I\ls S$^{(1)}$\ns}
\affiliation{
$^{(1)}$ Universit\'e de Lyon, Laboratoire de Physique de l'\'Ecole Normale Sup\'erieure de Lyon, CNRS, France\\
$^{(2)}$ Laboratoire M2P2, UMR 6181 CNRS-Universit\'{e}s Aix-Marseille, France\\
$^{(3)}$ Department of Mechanical Engineering, MIT, Cambridge, MA 01239 USA \\
$^{(4)}$ Laboratoire des \'{E}coulements G\'{e}ophysiques et Industriels (LEGI), CNRS, Grenoble, France.}

\date{\today}

\begin{document}

\maketitle

\begin{abstract}
We present the results of a combined experimental and numerical study of the generation of internal waves using the novel internal wave generator design of Gostiaux {\em et al.} (2007). This mechanism, which involves a tunable source comprised of oscillating plates, has so far been used for a few fundamental studies of internal waves, but its full potential has yet to be realized. Our studies reveal that this approach is capable of producing a wide variety of two-dimensional wave fields, including plane waves, wave beams and discrete vertical modes in finite-depth stratifications. The effects of discretization by a finite number of plates, forcing amplitude and angle of propagation are investigated, and it is found that the method is remarkably efficient at generating a complete wave field despite forcing only one velocity component in a controllable manner. We furthermore find that the nature of the radiated wave field is well predicted using Fourier transforms of the spatial structure of the wave generator.
\end{abstract}

\section{Introduction}

The study of internal waves continues to generate great interest due
to the evolving appreciation of their role in many geophysical
systems.  In the ocean, internal waves play an important role in
dissipating barotropic tidal energy \cite[see][for
a review]{bib:GarrettKunze07}, whereas atmospheric internal waves are
an important means of momentum transport
\cite*[][]{bib:Alexander06}. In both the ocean and the atmosphere,
internal wave activity also impacts modern day technology
\cite*[][]{bib:Osborne78}.  Many unanswered questions remain, however, particularly regarding the fate of internal waves. For example: How much mixing do internal waves
generate in the ocean, and via what processes? And at what altitudes do atmospheric internal waves break and deposit their momentum? The ability to reliably model internal
wave dynamics is key to tackling important questions such as these.

Internal waves come in a wide variety of forms.  The simplest, a plane wave, is the basis of many theoretical studies that provide fundamental insight \cite[see][for instance]
{bib:Thorpe87,bib:Thorpe98,bib:DauxoisYoung99}, especially as any linear wave structure can be decomposed into independent plane waves via Fourier transforms.
While plane wave solutions are the focus of many theoretical studies, laboratory experiments and field observations reveal that internal waves generated by a localized source, such as the tidal flow past an ocean ridge \cite*[][]{bib:Bell75,bib:Martinetal06} or deep tropical convection in the atmosphere \cite*[][]{bib:Walterscheidetal01}, produce coherent wave beams that radiate away from the generation site. In vertically-finite domains, such as the ocean, the internal wave field can be conveniently described using vertical modes, i.e. horizontally propagating and vertically standing waves, whose spatial form is dictated by the vertical stratification, as discussed in \cite{bib:Echeverrietal09}. Finally, localized solitary wave structures are also ubiquitous \cite[]{bib:NewDaSilva02}.

Investigation of these different internal wave forms in laboratory experiments has played a key role in internal wave research, starting with the pioneering work of \cite{bib:MowbrayRarity67} on the
wave beams generated by an oscillating cylinder. Since then, internal waves have been generated using a variety of means. \cite{bib:DelisiOrlanski75}
performed an experimental study of the reflection of nominally plane waves,
produced by a paddle mechanism, from a density jump. A similar paddle mechanism was used by \cite*{bib:Iveyetal00}
to study the dissipation caused by internal wave breaking at a sloping boundary. \cite{bib:Maasetal97} used vertical oscillations of a tank filled with salt-stratified water to parametrically excite
internal waves, which eventually focused onto internal wave attractors in the tank. With the ocean in mind, \cite{bib:GostiauxDauxois07} and \cite{bib:Echeverrietal09} produced internal waves by side-to-side oscillation of
topography.

The aforementioned experimental methods of internal wave generation have three inherent shortcomings. Firstly, they produce wave fields that are invariant in one horizontal direction, and thus nominally two-dimensional. In this paper, we too restrict ourselves to the study of such situations, using $z$ to refer to the vertical direction, antiparallel to the gravity field $\bm{g}=-g\bm{e}_{z}$, and $x$ to the horizontal direction; the possibility of generating three-dimensional wave fields using the novel generator is raised at the end of the paper. The second shortcoming is that, with the exception of towed topography \cite[][]{bib:Baines85,bib:Aguilar06}, pre-existing methods radiate waves in multiple directions rather than in a single direction. This is due to the dispersion relation for internal gravity waves,
\begin{equation}
\omega^2=N^2\sin^2\theta\,,
\label{eq:dispersion_relation}
\end{equation}
which relates the forcing frequency, $\omega$, to the local angle of energy propagation with respect to the horizontal, $\theta$, via the Brunt-V\"{a}is\"{a}l\"{a} frequency, $N=\sqrt{-g\partial_{z}\rho/\rho}$, where $\rho$ is the background fluid density. Since waves propagating
at angles $\pm\theta$ and $\upi\pm\theta$ all satisfy
(\ref{eq:dispersion_relation}) for a given frequency ratio $\omega/N$,
a two-dimensional localized source, such as a vertically oscillating
cylinder, generates internal waves propagating in four different
directions. Propagation in two of the four directions can be suppressed by either providing oscillations along only one of the directions of propagation (i.e. $\theta$ and $\upi+\theta$)
\cite[][]{bib:GavrilovErmanyuk96,bib:ErmanyukGavrilov08} or by using a paddle system at a boundary \cite[][]{bib:DelisiOrlanski75}. These arrangements nevertheless still produce an undesirable second set of waves that must somehow be dealt with in an experiment. The third, and perhaps the most significant, shortcoming is that all pre-existing methods provide very limited, if any, control of the spatial structure of an internal wave field.

\begin{figure}
\begin{center}
\begin{picture}(10.0,5)
\put(0.0,-0.3){\includegraphics[width=9.5cm]{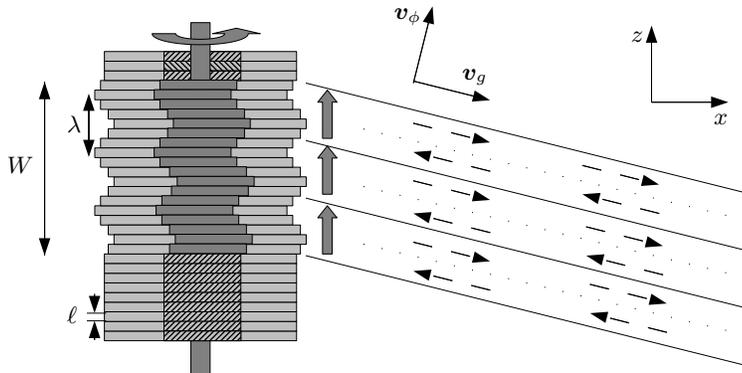}}
\put(-.4,2.45){$W$}
\put(.4,0.45){$\ell$}
\put(.4,3){$\lambda$}
\put(7.9,4.2){$z$}
\put(9,3.1){$x$}
\put(4.75,4.5){${\bm{v}_{\phi}} $}
\put(5.65,3.7){${\bm{v}_{g}}$}
\end{picture}
\end{center}
\caption{A schematic showing the basic configuration of a novel internal wave generator. Plates are vertically stacked on an eccentric camshaft.  See text in section 2 for the definitions of the different lengths $W$, $\ell$ and $\lambda$. The circular arrow at the top of the generator illustrates the direction of rotation of the camshaft, the thick vertical arrows show the corresponding motion of the wave form of the plates, and the dashed oblique arrows indicate the resulting local velocity field. ${\bm{v}_{\phi}} $ and ${\bm{v}_{g}}$ indicate the direction of phase and group velocity, respectively.}
\label{fig:principle_generator}
\end{figure}

A major advance in internal wave generation recently occurred with the design of a novel type of internal wave generator \cite[][]{bib:Gostiauxetal07}. This design uses a series of stacked, offset plates on a camshaft to simultaneously shape the spatial structure of an experimental internal wave field and enforce wave propagation in a single direction, as illustrated in figure \ref{fig:principle_generator}. The maximum horizontal displacement of each plate is set by the eccentricity of the corresponding cam, and the spatio-temporal evolution is defined by the
phase progression from one cam to another and the rotation speed of the camshaft. So far, this novel configuration has been used to study plane wave reflection from sloping
boundaries \cite[][]{bib:Gostiaux06}, diffraction through a slit \cite[][]{bib:Mercieretal08} and wave beam propagation through nonuniform stratifications \cite[][]{bib:MathurPeacock09}.  Despite these early successes, however, there has been no dedicated study of the ability of this arrangement to generate qualitatively different forms of
internal wave fields, and several important questions remain. For example, how does a stratified fluid that supports two-dimensional waves respond to controlled forcing in only one direction (i.e. parallel to the motion of the plates)?

In this paper, we present the results of a comprehensive study of two-dimensional wave fields produced by different configurations of novel internal wave generators, and reveal that this approach can accurately produce plane waves, wave beams and discrete vertical modes. The results of experiments are compared with predictions based on the Fourier transforms of the spatial structure of the wave generator, which proves to be a very useful and simple tool for predicting wave fields, and numerical simulations, which allow investigation of the boundary conditions imposed by the generator. The material is organized as follows. Section~\ref{sct:principles} presents the experimental and numerical methods used throughout the study. The generation of plane waves is addressed in \S~\ref{sct:planewaves}, followed by the generation of self-similar wave beams and vertical modes in \S~\ref{sct:stevenson} and \S~\ref{sct:mode}, respectively. Our conclusions, and suggestions for future applications of the generator, are presented in \S~\ref{sct:conclusion}.

\section{Methods}
\label{sct:principles}

\subsection{Experiments}
\label{subsct:principles}

Throughout this paper, we consider the case of wave fields excited by a vertically standing generator with horizontally moving plates of thickness $\ell$, as depicted in figure~\ref{fig:principle_generator}. This scenario, which is possible because the direction of wave propagation is set by the dispersion relation (\ref{eq:dispersion_relation}), has two major advantages over the other possibility of a generator tilted in the direction of wave propagation \cite[][] {bib:Gostiauxetal07,bib:MathurPeacock09}. First, it is far more convenient because it requires no mechanical components to orient the camshaft axis and no change of orientation for different propagation angles. Second, unwanted wave beams that are inevitably produced by free corners within the body of a stratified fluid are eliminated because the generator extends over the entire working height of the fluid.

Two different laboratory experimental facilities, both using
the double-bucket method \cite[][]{bib:Oster65} to create salt/density stratifications, were used.
The first, at ENS de Lyon, utilized a $0.8$~m long, $0.170$~m wide and $0.425$~m deep wave tank. The wave generator, whose characteristics are listed in table~\ref{table:generator}, was positioned at one end of the tank. On each side of the wave generator there was a $0.015$~m gap between the moving plates and the side wall of the wave tank. Visualizations and quantitative measurements of the density-gradient perturbation field were performed using the Synthetic Schlieren technique \cite[][]{bib:Dalzieletal00}. The CIV algorithm of \cite{bib:Fincham00} was used to compute the cross-correlation between the real-time and the $t=0$ background images. Blocksom filter matting was used to effectively damp end wall reflections of internal waves.  The ENS Lyon setup was used to run
experimental studies of the classical Thomas--Stevenson wave beam
profile \cite[][]{bib:ThomasStevenson72}, detailed in \S~\ref{sct:stevenson}.

The second system, at MIT, utilized a $5.5$~m long, $0.5$~m wide and $0.6$~m deep wave tank. A partition divided almost the entire length of the tank into $0.35$~m and $0.15$~m wide sections, the experiments being performed in the wider section. The wave generator, whose characteristics are given in table~\ref{table:generator}, was mounted in the $0.35$~m wide section of the tank with a gap of $0.025$~m between the moving plates and either side wall. Parabolic end walls at the ends of the wave tank reflected the wave field produced by the generator into the $0.15$~m wide section of the tank, where it was dissipated by Blocksom filter matting. Visualizations and quantitative measurements of the velocity field in the vertical midplane of the generator were obtained using a LaVision Particle Image Velocimetry (PIV) system. This facility was used for studies of plane waves and vertical modes, detailed in \S~\ref{sct:planewaves} and \S~\ref{sct:mode}, respectively.

\begin{table}
\centering
\begin{tabular}{cc @{.} ll @{.} ll @{.} ll @{.} ll @{.} ll}
Generator & \multicolumn{2}{c}{Height} & \multicolumn{2}{c}{Width} &
\multicolumn{2}{c}{Plate thick.} & \multicolumn{2}{c}{Plate gap}
 & \multicolumn{2}{c}{Max. eccentricity} & Num. of plates\\[3pt]
Lyon & 390&$0$ & $140$&$0$ & $6$&$0$ & $0$&$7$ & $10$&$0$ & $60$\\
MIT & $534$&$0$ & $300$&$0$ & $6$&$3$ & $0$&$225$ & $35$&$0$ & $82$
\end{tabular}
\caption{Details of the wave generators at ENS de Lyon and MIT. Dimensions are in mm.}
\label{table:generator}
\end{table}

\begin{figure}
\begin{center}
\begin{picture}(12.0,7.2)
\put(0.0,.25){\includegraphics[width=12.5cm]{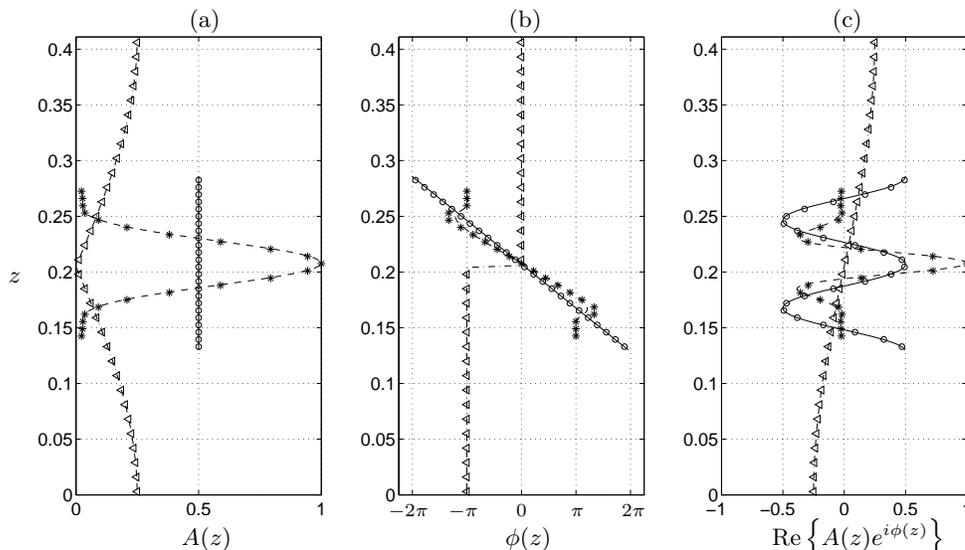}}
\put(2,6.75){(a)}
\put(6.25,6.75){(b)}
\put(10.55,6.75){(c)}
\put(-0.4,3.35){$z$}
\put(1.9,-0.15){$A(z)$}
\put(4.65,0.235){\scriptsize{$-2\pi$}}
\put(5.45,0.235){\scriptsize{$-\pi$}}
\put(6.35,0.235){\scriptsize{$0$}}
\put(7.075,0.235){\scriptsize{$\pi$}}
\put(7.75,0.235){\scriptsize{$2\pi$}}
\put(6.2,-0.15){$\phi(z)$}
\put(9.75,-0.15){$\Re\left\{A(z)e^{i\phi(z)}\right\}$}
\end{picture}
\end{center}
\caption{Examples of the (a) eccentricity $A(z)$ in cm, (b) phase $\phi(z)$ and (c) instantaneous position of the cams for different profiles used throughout the paper. These include: plane waves for $M=12$ and $W=2\lambda$ ($\circ$), Thomas $\&$ Stevenson beam ($\ast$) and a mode-1 internal tide ($\triangleleft$). Thin lines drawn through the discrete points are the corresponding analytical forms being modeled.}
\label{fig:excentricityandcames}
\end{figure}

Examples of the amplitude and phase arrangements of the plates for the experiments discussed in this paper are presented in
figure~\ref{fig:excentricityandcames}. We use the following terminology: $M$ is the number of plates per period used to represent a periodic wave form of vertical wavelength $\lambda$, $W$ is the total height of the active region of the generator with nonzero forcing amplitude, $A(z)$ is the eccentricity of a cam located at height $z$, and $\phi(z)$ is the phase of a cam set by the initial rotational orientation relative to the mid-depth cam ($\phi=0$). The actual profile of the generator is given by $\Re\left\{A(z)e^{i\phi(z)}\right\}$, where Re stands for the real part. For a plane wave, $A(z)$ is constant and $\phi(z)$ varies linearly over the active region of the generator. For the Thomas \& Stevenson profile there is a nontrivial spatial variation in both $A(z)$ and $\phi(z)$ over the active region, and elsewhere $A(z)$ is zero. Finally, for a mode-1 wave field, $A(z)$ varies as the magnitude of a cosine over the entire fluid depth, while $\phi(z)$ jumps by  $\upi$ at mid-depth.

\subsection{Numerics}
\label{numericalsimulation}

Complementary two-dimensional numerical simulations, in which excitation by the generator was modeled by imposing spatio-temporal variations of the velocity and buoyancy fields along one boundary of the numerical domain, were performed.  The simulations, which assumed a Newtonian fluid in the Boussinesq approximation, solved the incompressible continuity, Navier--Stokes and energy equations
\begin{subequations}
\begin{equation}
\bm{\nabla\cdot v}=0,
\label{eq:continuity}
\end{equation}
\begin{equation}
\partial_{t}\bm{v}+\left(\bm{\nabla}\times\bm{v}\right)\times\bm{v}
=-\bm{\nabla}q+b\bm{e}_{z}+\nu\nabla^{2}\bm{v}
\label{eq:Navier_Stokes}
\end{equation}
and
\begin{equation}
\partial_{t}b+\left(\bm{v\cdot\nabla}\right)b=-N^{2}\bm{v\cdot e}_{z}+\kappa\nabla^{2}b,
\label{eq:energy}
\end{equation}
\label{eq:master_equations}
\end{subequations}
\noindent\/where $\bm{v}=\left(u,w\right)$ is the velocity field, $v$ the corresponding velocity magnitude, $p=q-v^{2}/2\,$ the pressure, $b$ the buoyancy field, related to the density by $\rho=\rho_{0}\left(1-g^{-1}N^{2}z-g^{-1}b\right)$ where $\rho_{0}$ is the density at $z=0$, $\nu$ the kinematic viscosity and $\kappa$ the diffusivity.

The code used was an extension of that developed for channel flow \cite[][]{bib:Gilbert88}, to which the integration of the energy equation and the possibility of spatially varying and time
evolving boundary conditions on the plates were added,
as already presented in \cite{bib:Martinandetal06} for thermal
convection. The method proceeds as follows. A numerical solution in a rectangular domain
$[0,l_{x}]\times[0,l_{z}]$ is obtained by a tau-collocation
pseudo-spectral method in space, using Fourier modes in the
$z$-direction and Chebyshev polynomials in the confined
$x$-direction. This method very precisely accounts for the dissipative
terms.  The nonlinear and diffusion terms are discretized in
time by Adams--Bashforth and Crank--Nicolson schemes, respectively,
resulting in second order accuracy in time.  As the simulation focuses
on the linear and weakly nonlinear dynamics, the de-aliasing of the
nonlinear term in the spatial expansion of the solution is not of
crucial importance and is omitted to decrease the computational
cost.  Owing to the assumption of a divergence-free flow, the
influence matrix method, introduced in
\cite{bib:KleiserSchumann80,bib:KleiserSchumann84}, is used to evaluate
the pressure and the velocity field; the pressure gradient
in the Navier--Stokes equation then being discretized by an implicit
Euler scheme.  Finally, the buoyancy term is also discretized in time
by an implicit Euler scheme, since the energy equation is solved before
the Navier--Stokes equations.

Simulations were run by imposing forced boundary conditions on components of the velocity and buoyancy fields at $x=0$; no forcing was applied to the pressure, since its value on the boundaries is an outcome of the numerical method.
The governing equations~(\ref{eq:master_equations}) were thus integrated together with Dirichlet boundary conditions
\begin{equation}
\bm{v}\left(0,z,t\right)=\bm{v}_{f}\left(z,t\right),\,
b\left(0,z,t\right)=b_{f}\left(z,t\right);
\label{eq:topboundcond}
\end{equation}
while
\begin{equation}
\bm{v}\left(l_{x},z,t\right)=\bm{0},\,b\left(l_{x},z,t\right)=0,
\label{eq:bottomboundcond}
\end{equation}
were applied at $x=l_{x}$. We note that this numerical forcing is Eulerian in nature, whereas the corresponding experimental forcing is Lagrangian in spirit. The spectral method introduces periodic conditions in the $z$-direction, which have to be accounted for to avoid Gibbs oscillations.
Therefore, the boundary conditions (\ref{eq:topboundcond}) were multiplied by a polynomial ``hat'' function $H\left(z\right)=\left(1-\left(2z/l_{z}-1\right)^{30}\right)^{6}$, vanishing at $z=0$ and $z=l_{z}$. The choice of the exponents in $H\left(z\right)$ is qualitative, the aim being that the variation of the profile envelope be smooth compared to the spatial resolution, yet sharp enough to keep a well-defined width of forcing.

The boundary conditions (\ref{eq:bottomboundcond}) imply wave reflection, with the reflected waves eventually interfering with the forced waves. Thus, the numerical domain was made sufficiently large to establish the time-periodic forced wave field near the generation location long before reflections became an issue. For a typical simulation, the domain was $l_x=3.01$~m long and $l_z=1.505$~m high, and the number of grid points used was $N_{x}=1024$ and $N_{z}=512$, giving a spatial vertical resolution of $2.9$~mm that ensured at least two grid points per plate. Satisfactory spectral convergence was confirmed for this spatial resolution and the time step was set to ensure stability of  the numerical scheme.

\subsection{Analysis}
\label{subsct:fouranal}

A detailed study of the impact of sidewall boundary conditions on the generation of shear waves was performed by \cite{bib:McEwan74}. Here, we take a simpler approach and show that a useful tool for investigating both theoretical
and experimental internal wave fields produced by the novel generator is Fourier analysis. This allows one to decompose
internal wave fields into constituent plane waves, and readily make predictions about the radiated
wave field.

For an unconfined, inviscid, two-dimensional system, any physical field variable associated with a periodic internal wave field of frequency $\omega$  can be described by its Fourier
spectrum~\cite*[][]{bib:TabaeiAkylas03,bib:Tabaeietal05}, i.e.
\begin{equation}
\psi(x,z,t)= \frac{e^{-\mathrm{i}\omega t}}{2\upi}
\int_{-\infty}^{+\infty}\!\!\!\!\int_{-\infty}^{+\infty}\!\!\!\!
\,\tilde{Q}_{\psi}(k_x,k_z)e^{\mathrm{i}(k_x x+k_z
z)}\,\delta\left((k_x^2+k_z^2)\omega^2-N^2 k_x^2\right)\,\mbox{d}k_z\mbox{d}k_x,
\label{psi_general}
\end{equation}
where $\psi(x,z,t)$ represents a field variable (e.g. $b$, $u$) and the Dirac $\delta$-function ensures the dispersion relation~(\ref{eq:dispersion_relation}) is satisfied by the plane waves components. Propagating waves in a single direction, say towards positive $x$-component and negative $z$-component for the energy propagation, require
\begin{equation}
\tilde{Q}_{\psi}(k_x,k_z)=0 \quad\forall k_x \leq 0 \quad \& \quad \forall
k_z \leq 0,
\label{eq:criteria}
\end{equation}
as noted by \cite{bib:Mercieretal08}.

At a fixed horizontal location $x_0$, the values of $\psi(x_0,z,t)$ for all the field variables can be considered as boundary conditions that force the propagating wave field $\psi(x,z,t)$.
Knowing the Fourier transform of the boundary forcing,
\begin{equation}
Q_{\psi}(x_{0},k_z) = \frac{e^{\mathrm{i}\omega t}}{\sqrt{2\upi}}\int_{-\infty}^{+\infty}\,
\psi(x_{0},z,t)\,e^{-\mathrm{i}k_{z}z}\,\mbox{d}z,
\label{def_Qz}
\end{equation}
leads to complete description of the radiated wave field for $x>x_{0}$:
\begin{equation}
\psi(x,z,t)=  \frac{e^{-\mathrm{i}\omega t}}{2\upi}\int_{0}^{+\infty} \!\!\!\!
\int_{-\infty}^{+\infty} \!\!\!\!
Q_{\psi}(x_{0},k_z)\,e^{\mathrm{i}(k_xx + k_zz)}\delta\left((k_x^2+k_z^2)\omega^2-N^2 k_x^2\right)\,
\mbox{d}k_z\,\mbox{d}k_x,
\label{eq:def_Qx}
\end{equation}
assuming that only right-propagating waves (i.e. $k_x\geq0$) are possible.

In practice, the novel wave generator we consider forces only the horizontal velocity field in a controlled manner, i.e.
\begin{equation}
\psi(0,z,t)=u(0,z,t)=\mathrm{Re}\left\{U(z)e^{-i\omega t}\right\}.
\label{eq:bc}
\end{equation}
As such, we expect the Fourier transform of this boundary condition to act only as a guide for the nature of the radiated wave field, since it is not clear how the fluid will respond to forcing of a single field variable. Throughout the paper, we perform the Fourier transform along a specific direction using the Fast Fourier Transform algorithm. To compare spectra from theoretical, numerical and experimental profiles with the same resolution, a cubic interpolation (in space) of the experimental wave field is used if needed.

Unless otherwise stated, the experimental and numerical results presented are filtered in time at the forcing frequency $\omega$. The aim is to consider harmonic (in time) internal waves for which we can define the Fourier decomposition in (\ref{psi_general}), and to improve the signal-to-noise ratio, which lies in the range $10^{2}-10^{1}$, with the best results for shallow beam angles and small amplitude forcing. The time window $\Delta t$ used for the filtering is such that $\omega\Delta t/2\pi\geq 9$ for $A_0=0.005$~m and $\omega\Delta t/2\pi\geq4$ for $A_0=0.035$~m (where $A_0$ is the amplitude of motion of the plates defined in \ref{geneconsiderations}), ensuring sufficient resolution in Fourier space for selective filtering. The recording was initiated at time $t_0$ after the start-up of the generator such that $Nt_0/2\pi\simeq30\gg1$, ensuring no transients remained. 
~\cite[][]{bib:Voisin03}.

\section{Plane waves}
\label{sct:planewaves}

Since many theoretical results for internal waves are obtained for plane waves \cite[e.g.][]{bib:Thorpe87,bib:Thorpe98,bib:DauxoisYoung99}, the
ability to generate a good approximation of a plane wave in a laboratory setting is important to enable corresponding experimental investigations. In order to generate a nominally plane wave, however, one must consider the impact of the different physical constraints of the wave generator, which include: the controlled forcing of only one velocity component by the moving plates, the finite spatial extent of forcing, the discretization of forcing by a finite number of plates, the amplitude of forcing, and the direction of wave propagation with respect to the camshaft axis of the generator. In this section, we present the results of a systematic study of the consequences of these constraints; a summary of the experiments is presented in table~\ref{table:exp}.

\subsection{Analysis}
\label{geneconsiderations}

Two-dimensional, planar internal waves take the form $\psi\left(x,z,t\right)=\Re\left\{\psi_{0}e^{\left(\mathi k_{x}x+\mathi k_{z}z-\mathi\omega t\right)}\right\},$ where $\bm{k}=\left(k_{x},k_{z}\right)$ is the wave vector and $\psi(x,z,t)$ represents a field variable.
Being of infinite extent is an idealization that is never realizable in an experiment. To investigate the consequences of an internal wave generator being of finite extent, the horizontal velocity boundary conditions used to produce a downward, right-propagating, nominally plane wave can be written as:
\begin{equation}
u(0,z,t) = \mathrm{Re}\left\{\left[ \Theta(z+W/2)-\Theta(z-W/2)
\right]\ A_0\omega\ e^{i(-\omega t + k_e z)}\right\}\,,
\label{eq:bc1}
\end{equation}
where $k_e>0$ is the desired vertical wave number, $\lambda_{e}=2\upi/k_{e}$ the corresponding vertical wavelength,
$-W/2\leq\,z\leq\,W/2$ the vertical domain over which forcing is applied, $A_0$ the amplitude of motion
of the plates, and $\Theta$ the Heaviside function. The spatial Fourier transform of
(\ref{eq:bc1}) is:
\begin{equation}
Q_{u}(0,k_z)=A_0\omega \frac{W}{\sqrt{2\upi}}\mathrm{sinc}\left(\frac{(k_z-k_e)W}{2}\right)\,,
\label{eq:planewave_ft}
\end{equation}
with $\mathrm{sinc}(x)=\sin x/x$ the sine cardinal function. In the limit $W\rightarrow\infty$, Eq.~(\ref{eq:planewave_ft}) approaches a delta function, which is the Fourier transform of a plane wave.
Owing to the finite value of $W$, however, $Q_{u}(0,k_z)$ does not vanish
for negative values of $k_z$, suggesting that (\ref{eq:bc1}) will also excite upward propagating plane waves.
Following the convention usual in optics that $Q_{u}(0,k_z)$ is
negligible for $|k_z-k_e|W/2\geq \upi$, if $k_e\gg 2\upi/W$ then~(\ref{eq:criteria}) is reasonably satisfied.

Another consideration is that the forcing provided by the wave generator is not spatially continuous, but discretized by $N_p$ oscillating plates of width $\ell$. Accounting for this, the boundary forcing can be written as:
\begin{equation}
u(0,z,t) = \Re\left\{\sum_{j=0}^{N_p-1} \left[
\Theta(z-z_{j})-\Theta(z-z_{j}+\ell) \right] A(z_j)\, e^{i(k_e z_j + k_e\ell/2 -\omega t )}\right\}\,,
\label{eq:forcing2}
\end{equation}
where $z_j=j\ell-W/2$ and $N_p\ell=W$. The Fourier transform of (\ref{eq:forcing2}) for the specific case of constant amplitudes $A(z_j)=A_{0}\omega,~\forall j\in\{0,\cdots,N_p-1\}$, reduces to
\begin{equation}
Q_{u}(0,k_z)=A_{0}\omega\frac{\ell}{\sqrt{2\upi}}\mathrm{sinc}\left(\frac{k_z\ell}{2}\right)
\frac{\displaystyle\sin\left(\frac{(k_z-k_e)W}{2}\right)}
{\displaystyle\sin\left(\frac{(k_z-k_e)\ell}{2}\right)}\,,
\label{eq:Qdiscrete_Aconstant}
\end{equation}
a classical result often encountered for diffraction gratings.
Consequently, the magnitudes of $W$ and $\ell$ in comparison to the desired vertical wavelength $\lambda_{e}={2\upi}/{k_{e}}$ characterize the spread of the Fourier spectrum and the potential for excitation of upward propagating waves.

In the following sections, we quantify the downward emission of waves using the parameter $\beta_{d}$, defined as:
\begin{equation}
\beta_{d}=\frac{\displaystyle\int_0^{+\infty}|Q_{u}(x_{0},k_z)|^{2}\,\mbox{d}k_z}{\displaystyle\int_{-\infty}^{+\infty}|Q_{u}(x_{0},k_z)|^{2}\,\mbox{d}k_z},
\label{eq:estimateparameter2}
\end{equation}
which is essentially the ratio of the total kinetic energy of the downward-propagating waves to the total kinetic energy of the radiated wave field.

\begin{table}
\centering
\begin{tabular}{cccccccc}
 & Expt. & Experiment/Simulation & Forcing & $M$ & $W/\lambda_{e}$ & $A_0$ (mm) & $\theta$ (deg.) \\
\hline
Common case & 1 & exp./sim. & partial & 12 & 3 & 5.0 & 15  \\
\hline Forcing & 2 & sim. & complete & 12 & 3 & 5.0 & 45  \\
\hline \multirow{4}{*}{Angle} & 3 & exp./sim. & partial & 12 & 3 & 5.0 & 30 \\
& 4 & exp./sim. & partial & 12 & 3 & 5.0 & 45 \\
& 5 & exp./sim. & partial & 12 & 3 & 5.0 & 60 \\
& 6 & sim. & partial & 12 & 3 & 5.0 & 75 \\
\hline \multirow{4}{*}{Width} & 7 & exp./sim. & partial & 12 & 2 & 5.0 & 15 \\
& 8 & exp./sim. & partial & 12 & 1 & 5.0 & 15 \\
& 9 & exp./sim. & partial & 12 & 2 & 5.0 & 45 \\
& 10 & exp./sim. & partial & 12 & 1 & 5.0 & 45 \\
\hline \multirow{2}{*}{Discretization} & 11 & sim. & partial & $\infty$ & 3 & 5.0 & 15  \\
& 12 & exp./sim. & partial & 4 & 3 & 5.0 & 15  \\
\hline \multirow{3}{*}{Amplitude} & 13 & exp./sim. & partial & 12 & 3 & 35.0 & 15 \\
& 14 & exp. & partial & 12 & 3 & 35.0 & 30 \\
& 15 & exp. & partial & 12 & 3 & 35.0 & 45 \\
\end{tabular}
\caption{Summary of experiments and numerical simulations. $M$ is the number of plates used for one wavelength, $W/\lambda_{e}$ is the spatial extent of forcing expressed in terms of the dominant wavelength, $A_0$ is the eccentricity of the cams, and $\theta$ is the energy propagation angle. For complete forcing, $u$, $w$ and $b$ were forced at the boundary.}
\label{table:exp}
\end{table}

\subsection{Configuration}

The MIT facility was used for these plane wave experiments  (see table~\ref{table:generator}). A variety of different configurations were tested and these are summarized in table \ref{table:exp}. The fluid depth was  $H=0.56\pm 0.015$~m and the background stratification was $N=0.85$~rad~s$^{-1}$ for all experiments. An example configuration of the cams (amplitude and phase evolution) is presented in figure~\ref{fig:excentricityandcames}.
Plane waves were produced by configuring the $N_p$ plates of the wave generator with an
oscillation amplitude $A_{0}= 0.005$~m, with the exception of experiments 13 to 15 for which $A_{0}= 0.035~m$.
Results were obtained for different forcing frequencies corresponding to propagating
angles of 15, 30, 45 and 60$^{\circ}$.
Visualization of the wave field was performed using PIV, for which it was possible
to observe the wave field in a 40~cm-wide horizontal domain over the entire depth of the tank, save for a 1 cm loss near the top and bottom boundaries due to unavoidable laser reflections. The corresponding numerical simulations were configured accordingly.

\subsection{Results}

\subsubsection{Forcing}
\label{subsct:horizontal}

The consequences of forcing only a single component of the velocity field, which we call {\it partial forcing}, in comparison to forcing both the velocity field components and the buoyancy field (assuming they are related by the inviscid linear wave equation), which we call {\it complete forcing}, were investigated first using the numerical simulations.
Here, we present the results of simulations performed using a sinusoidal boundary wave form with $W=3\lambda_e$, $M=12$ and $\lambda_{e}=78.8$~mm (experiments 2 and 4 of table~\ref{table:exp}). The magnitude and frequency of the boundary condition for horizontal velocity were $A_{0}=5.0$~mm and $\omega=0.601$~rad~s$^{-1}$, the latter giving $\theta=45^{\circ}$.

\begin{figure}
\begin{center}
\begin{picture}(12.0,5.5)
\put(-0.5,0.0){\includegraphics[height=5cm]{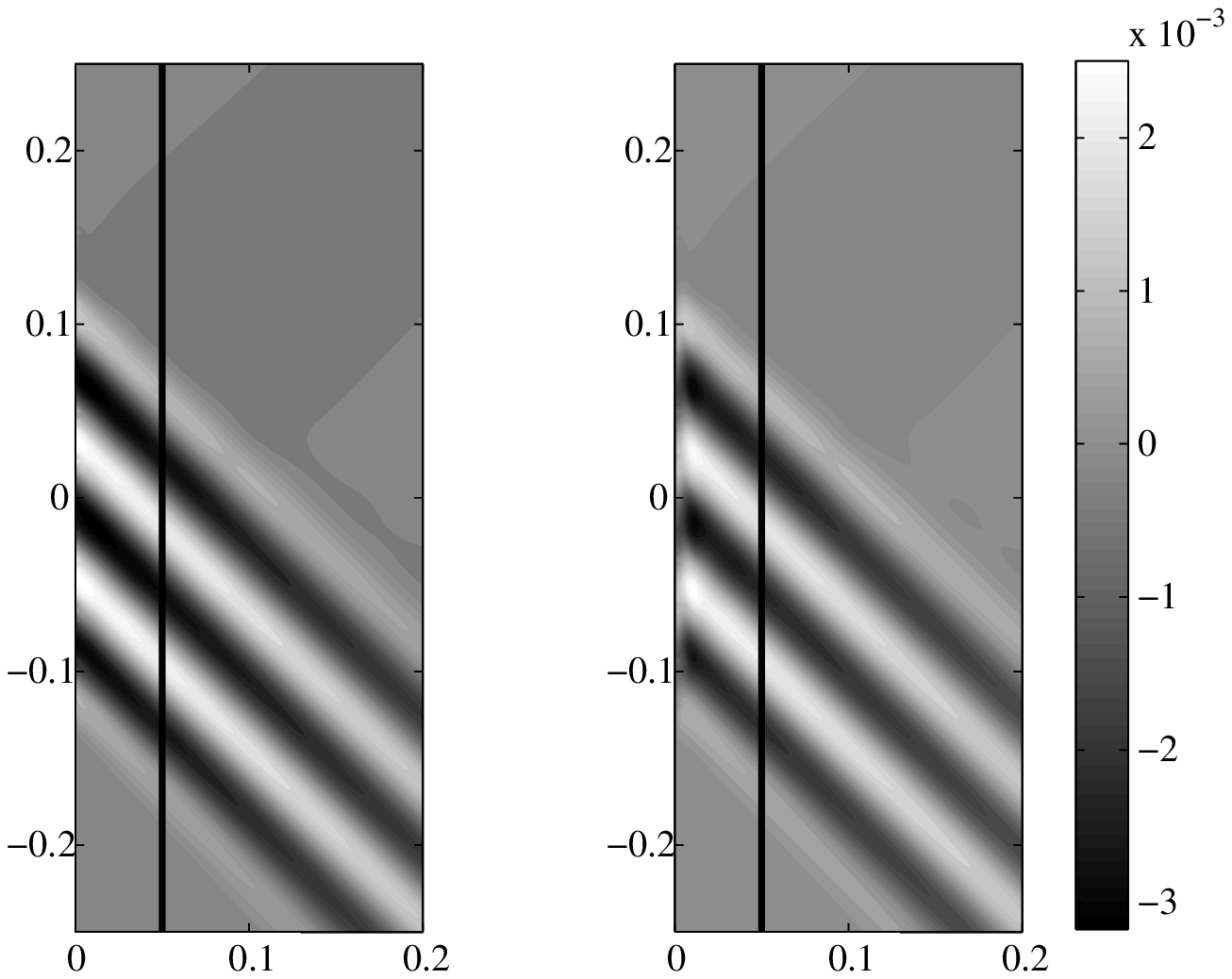}}
\put(6.6,0.0){\includegraphics[height=5cm]{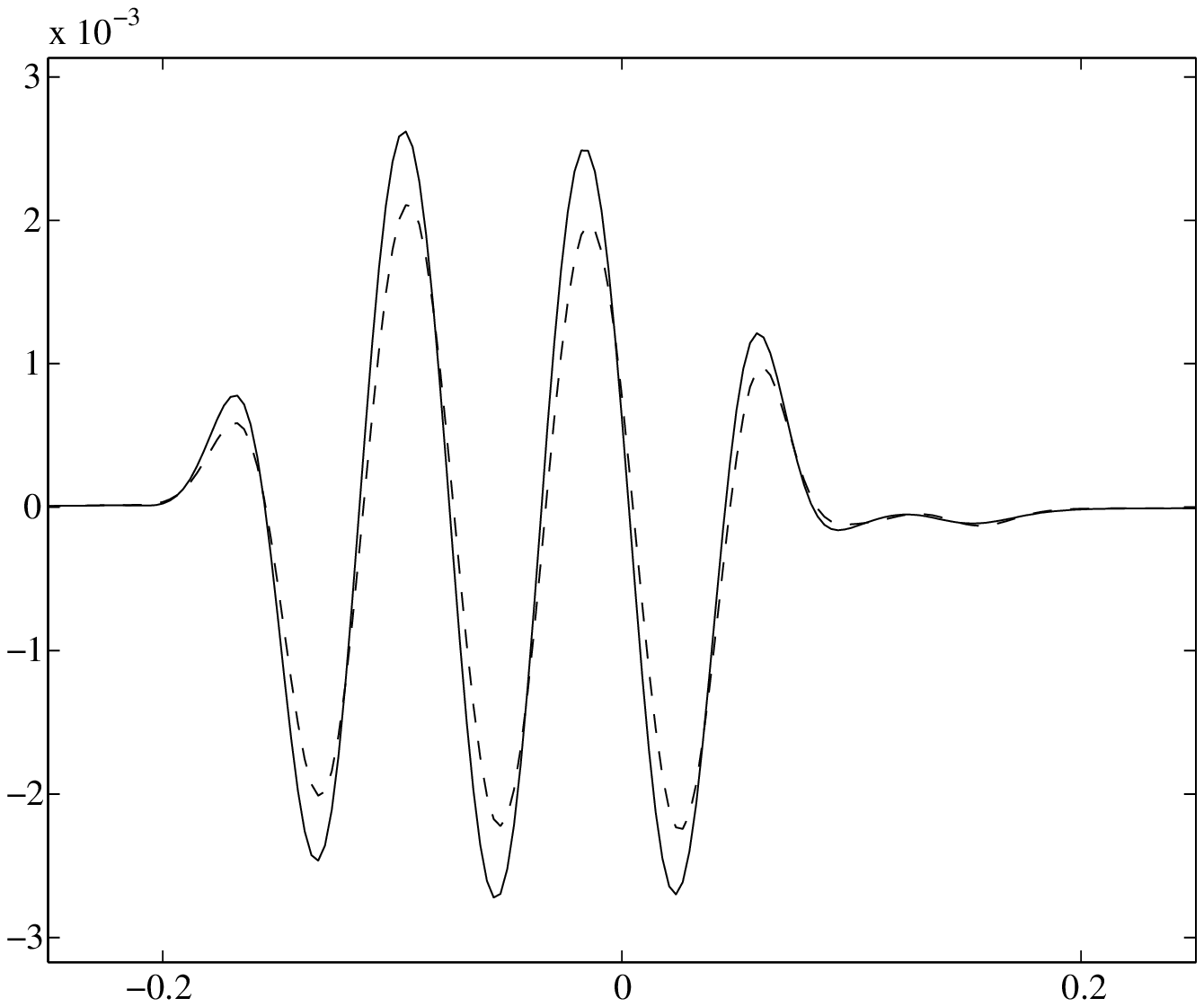}}
\put(0.6,4.9){(a)}
\put(3.7,4.9){(b)}
\put(9.5,4.9){(c)}
\put(0.7,-0.25){$x$}
\put(0.5,0.4){${\cal{C}}$}
\put(-0.6,2.425){$z$}
\put(3.8,-0.25){$x$}
\put(3.6,0.4){${\cal{C}}$}
\put(2.5,2.425){$z$}
\put(9.6,-0.25){$z$}
\put(6.5,5.1){$u\left(x_{\cal C},z\right)$}
\end{picture}
\end{center}
\caption{Comparison of the numerically obtained horizontal velocity field, $u$, for plane wave beams forced by (a) complete and (b) partial forcing (experiments 2 and 4 in table~\ref{table:exp}). (c) Horizontal velocity along the cut ${\cal{C}}$ in (a) and (b), located at $x_{\cal{C}}=0.05$~m for complete ($-$) and partial ($--$) forcing. All lengths are in m and all velocities in m~s$^{-1}$.}
\label{fig:comp_forcing}
\end{figure}

Figures \ref{fig:comp_forcing}(a) and (b) present snapshots of the horizontal velocity fields $u$ produced by partial and complete boundary forcing, respectively, and there is excellent qualitative agreement between the two. More quantitative comparisons are provided in figure \ref{fig:comp_forcing}(c), which presents data along the vertical cuts ${\cal C}$ indicated in figures \ref{fig:comp_forcing}(a) and (b), located at $x_{\cal C}=0.05$~m; this location was chosen because it is close enough to the generator that viscous damping has only an order 1\% effect on the wave field (based on the linear viscous theory for plane waves by \cite{bib:Lighthill1978}), yet is sufficiently far from the generator to allow the wave field to adapt to the boundary forcing.
The cross beam profiles in figure \ref{fig:comp_forcing}(c) show the amplitude of $u$ is roughly $20$\% lower for the case of partial forcing compared to complete forcing, but otherwise their forms closely match. Spectral information reveals no other discernable difference between the two, and both cases give $\beta_d > 0.99$, compared to the theoretical prediction of $\beta_d = 0.97$, revealing that almost all the energy is being emitted downward.
Qualitatively and quantitatively similar results to those presented in figures \ref{fig:comp_forcing}(c) and~(d) were obtained for different physical quantities (e.g. $w$ and $b$) of the wave fields, and for the other configurations of the generator listed in table \ref{table:exp}.

\subsubsection{Angle of emission}
\label{subsct:angle}

\begin{figure}
\begin{center}
\begin{picture}(13.0,5.5)
\put(0.0,0.0){\includegraphics[width=13.0cm]{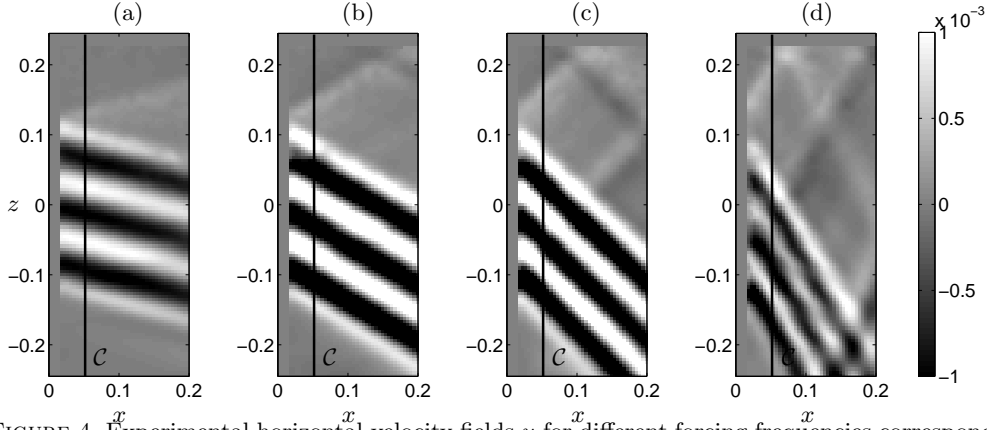}}
\put(1.4,5.05){(a)}
\put(4.45,5.05){(b)}
\put(7.5,5.05){(c)}
\put(10.55,5.05){(d)}
\put(1.4,-0.3){$x$}
\put(4.45,-0.3){$x$}
\put(7.5,-0.3){$x$}
\put(10.55,-0.3){$x$}
\put(0.0,2.5){$z$}
\put(1.15,0.45){${\cal{C}}$}
\put(4.2,0.45){${\cal{C}}$}
\put(7.25,0.45){${\cal{C}}$}
\put(10.3,0.45){${\cal{C}}$}
\end{picture}
\end{center}
\caption{Experimental horizontal velocity fields $u$ for different
forcing frequencies corresponding to angles of propagation
(a) $\theta=15^\circ$ , (b) $30^\circ$ , (c) $45^\circ$ and (d) $60^\circ$ . These are experiments 1, 3, 4 and 5 in table \ref{table:exp}, respectively.  All lengths are in m and all velocities in m~s$^{-1}$.}
\label{fig:different_angles_U}
\end{figure}

As one might expect, the quality of the wave field was best for shallower propagation angles, for which the horizontal velocity is a more defining quantity, and degraded for steep propagation angles, where $w$ becomes the dominant velocity component.
This is demonstrated in a qualitative manner by figures~\ref{fig:different_angles_U}(a)-(d), which present snapshots of the experimental horizontal velocity fields for $\theta =$ 15, 30, 45 and $60^\circ$, corresponding to experiments 1, 3, 4 and 5 in table \ref{table:exp}, respectively. Although the forcing velocity $u=A_{0}\omega$ is strongest for $\theta=60^\circ$, the wave field resulting from this partial forcing is not as strong and coherent as those at lesser angles.

\begin{figure}
\begin{center}
\begin{picture}(13.0,6.0)
\put(6.90,0.25){\includegraphics[width=6.40cm]{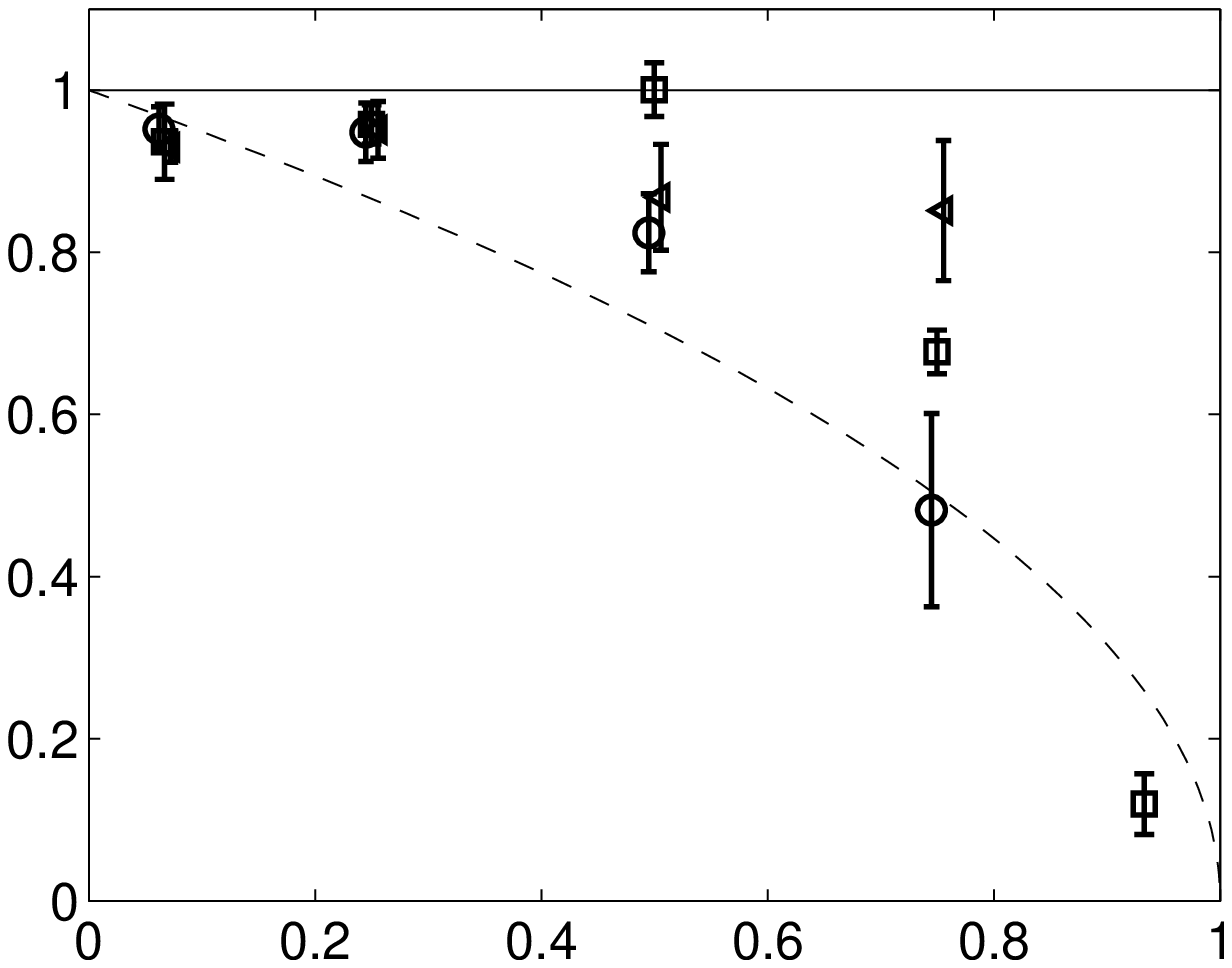}}
\put(-0.3,0.25){\includegraphics[width=6.40cm]{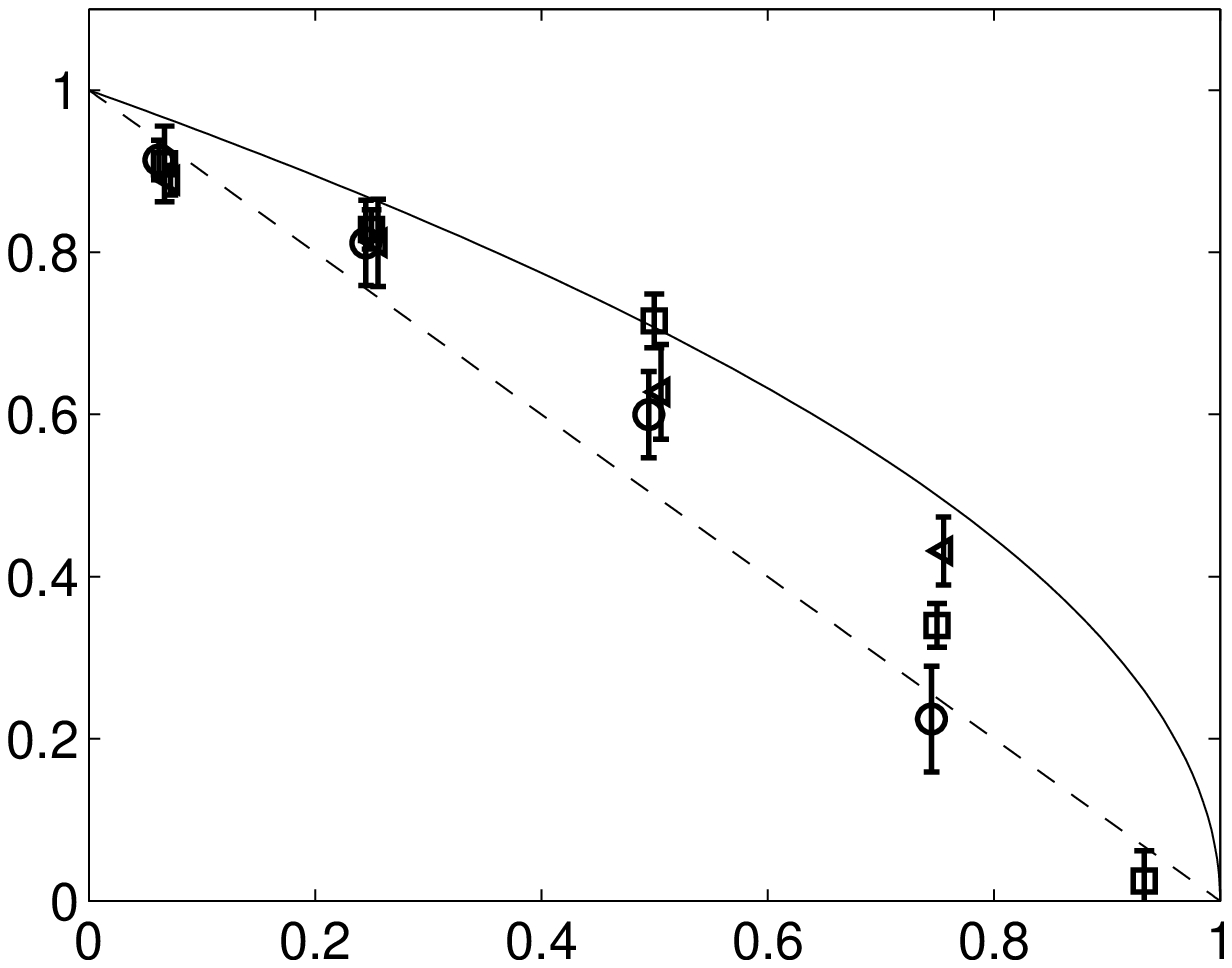}}
\put(-0.35,5.5){$u\left(x_{\cal{C}}\right)/A_{0}\omega$}
\put(6.75,5.5){$u^{\prime}\left(x_{\cal{C}}\right)/A_{0}\omega$}
\put(2.6,-0.25){$\left(\omega/N\right)^{2}$}
\put(9.8,-0.25){$\left(\omega/N\right)^{2}$}
\put(2.9,5.4){(a)}
\put(10.05,5.4){(b)}
\end{picture}
\end{center}
\caption{Comparison of numerical results ($\square$) and experimental results for the vertical ($\circ$) and tilted ($\triangleleft$) wave generator for the mean amplitude and standard deviation of (a) $u/A_{0}\omega$ and (b) $u^{\prime}/A_{0}\omega$. Results are plotted as a function of $(\omega/N)^2$ for different frequencies corresponding to $\theta=15$, $30$, $45$, $60$ and $75^\circ$ (numerics only). The solid lines correspond to $u^{\prime}/(A_{0}\omega)=1$ and the dashed lines to $u^{\prime}/(A_{0}\omega)=\cos\theta$.}
\label{fig:amplitude_beam}
\end{figure}

Another demonstration of the consequences of partial forcing is given in figure~\ref{fig:amplitude_beam}, which presents the efficiency of the wave generator as a function of the forcing frequency for experiments and numerics.
The efficiency is defined as the magnitude of the horizontal velocity component $u$ (figure~\ref{fig:amplitude_beam}(a)), or the velocity in the direction of wave propagation $u^{\prime}$ (figure~\ref{fig:amplitude_beam}(b)), averaged over the central $0.1$~m of the cut $\cal{C}$ across the wave field, compared to the forced horizontal velocity~$A_{0}\omega$. For complete forcing, one expects the ratio $u/A_{0}\omega$ to be one, whereas a smaller value indicates a less efficient mechanism. The results show that both $u/A_{0}\omega$ and $u^{\prime}/A_{0}\omega$ decrease with increasing $\omega/N$. A simple physical argument for the decay of the response with the propagation angle could be that the generator provides an initial amount of kinetic energy that is redistributed by the flow into both horizontal and vertical motions. If all the energy is appropriately distributed, we expect $u^{\prime}=A_{0}\omega$, and thus $u=A_{0}\omega\cos\theta=A_{0}\omega\sqrt{1-(\omega/N)^2}$, represented by solid lines in both panels of figure~\ref{fig:amplitude_beam}. Below $(\omega/N)^2=0.5$ the experiments and numerics follow this trend quite closely, but then depart from it significantly for higher frequency ratios. Also plotted as dashed lines in figure~\ref{fig:amplitude_beam} is the relation $u^{\prime}=A_{0}\omega\cos\theta$, implying $u^{\prime}/A_{0}\omega=\sqrt{1-\omega^2/N^2}$ and $u/A_{0}\omega={1-\omega^2/N^2}$, which does a reasonable job of capturing the trend of the results, especially at higher frequency ratios.
This relation implies that the energy associated with the motion of the plates along the direction of propagation is primarily responsible for setting the strength of the wave field.

Finally, we analyze the evolution the wave field with increasing angle by computing $\beta_d$. For these experiments, the Fourier analysis in section~\ref{geneconsiderations} predicts $\beta_d = 0.97$, independent of the angle of emission. Although the numerical values concur with this prediction, with $\beta_d>0.99$ for all angles, we obtain values of 0.99 for $\theta=15^{\circ}$, 0.98 for $\theta=30^{\circ}$ and $45^{\circ}$, and 0.93 for $\theta=60^{\circ}$ for the experiments.
One possible reason for this decrease in efficiency is the finite-amplitude lateral displacement of the plates of the generator, which could partially block the propagation of steeper waves; this is not taken into account in the numerics. Another possible reason is that at higher frequency of forcing the Reynolds number for the oscillating plates is larger, increasing the likelihood of more complex dynamics near the oscillating plates and thereby weakening their coupling to wave generation. These issues are raised again in section~\ref{subsct:amplitude}.


Overall, the results in this section and the previous section reveal that partial (horizontal) forcing by a vertically standing generator works well for $\theta\leq45^{\circ}$. This gives the user the freedom to perform experiments over a range of angles without having to re-orient the generator.
For larger angles, however, it would seem prudent to use a generator with its perpendicular axis tilted toward the direction of propagation. This was confirmed by a series of experiments similar to cases 1 and 3 to 5 in table~\ref{table:exp}, but with the generator tilted at $15^{\circ}$ to the vertical; these results are also included in figure~\ref{fig:amplitude_beam}. For angles smaller than $45^{\circ}$, there is almost no difference in the efficiency of the generator, but we found that the $60^{\circ}$ wave field produced by the tilted generator was noticeably stronger and more coherent than that produced by the vertically-standing generator.

\subsubsection{Finite extent}
\label{subsct:limitations}

The Fourier analysis in section~\ref{geneconsiderations} predicts that a consequence of a generator inevitably being of finite vertical extent is the production of undesirable waves that propagate in the vertical direction opposite to the principal wave field. This can be seen in both the numerical and experimental wave fields in figures~\ref{fig:comp_forcing} and \ref{fig:different_angles_U}, which contain a weak, upward-propagating wave field in addition to the principal  downward-propagating wave field. Experiments were therefore performed for $W/\lambda_e$ = 1, 2 and 3 to investigate how the strength of the undesirable wave field was influenced by the vertical restriction of the forcing, and to determine how well this was predicted by simple Fourier analysis. Except for varying $W/\lambda_e$ and having $\theta=15^{\circ}$ , the configuration was the same as in the previous subsection. The experiments are listed as experiments 1, 7 and 8 in table \ref{table:exp}.

\begin{figure}
\begin{center}
\begin{picture}(13.0,7.2)
\put(0,0){\includegraphics[width=13.cm]{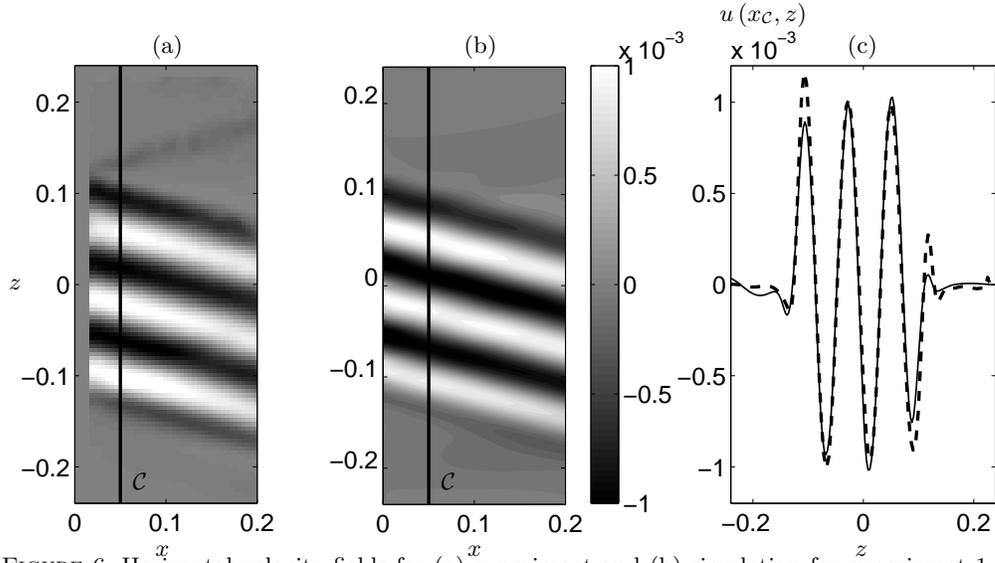}}
\put(1.75,6.4){(a)}
\put(5.9,6.4){(b)}
\put(11,6.4){(c)}
\put(1.8,-0.30){$x$}
\put(5.95,-0.30){$x$}
\put(-0.15,3.25){$z$}
\put(11.1,-0.30){$z$}
\put(1.5,0.6){${\cal {C}}$}
\put(5.6,0.6){${\cal {C}}$}
\put(9.3,6.825){$u\left(x_{\cal{C}},z\right)$}
\end{picture}
\end{center}
\caption{Horizontal velocity fields for (a) experiment and (b) simulation for experiment 1 in Table~\ref{table:exp}. (c) Vertical profiles along the cut $\cal{C}$ located at $x_{\cal {C}}=0.05$~m for the experiment ($--$) and the simulation ($-$). All lengths are in m and all velocities in m~s$^{-1}$.}
\label{fig:finiteextent}
\end{figure}

A direct comparison of experimental and numerical horizontal velocity fields for $W/\lambda_e=3$ is presented in figure~\ref{fig:finiteextent}, demonstrating very good agreement between the two, providing confirmation that our numerical approach can reliably model the horizontal forcing provided by the plates. The profiles presented in figure~\ref{fig:finiteextent}(c), obtained at the vertical cuts indicated in figures~\ref{fig:finiteextent}(a) and \ref{fig:finiteextent}(b), have only one small, but noticeable, difference: slightly higher peaks at either end of the experimental velocity profile. A similar level of agreement was obtained for the vertical velocity profile.

\begin{figure}
\begin{center}
\begin{picture}(10.0,9.2)
\put(0,0.25){\includegraphics[width=10.0cm]{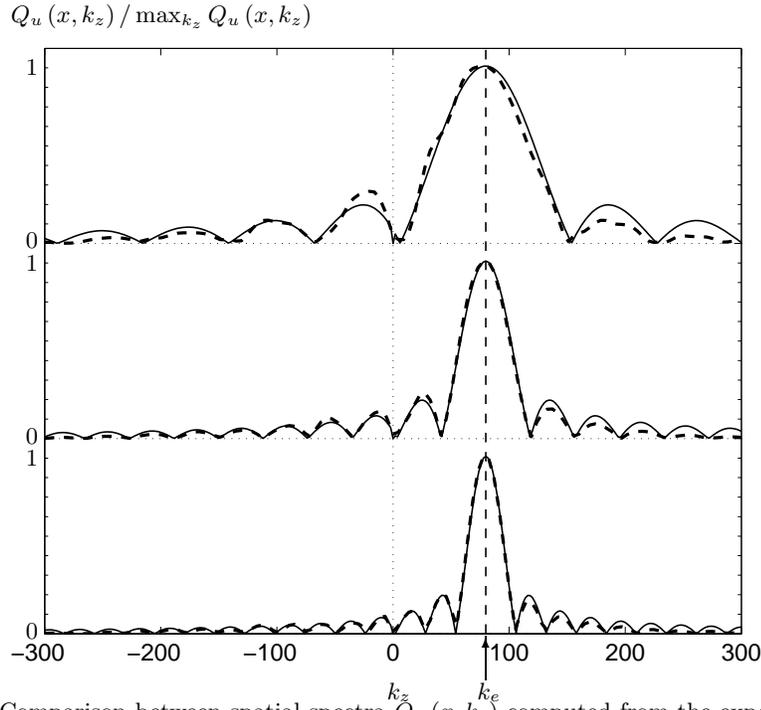}}
\put(5.,-0.25){$k_{z}$}
\put(6.2,-0.25){$k_{e}$}
\put(6.31,0){\vector(0,1){0.58}}
\put(0.0,8.75){$Q_{u}\left(x,k_{z}\right)/\max_{k_{z}}Q_{u}\left(x,k_{z}\right)$}
\put(0.2,0.55){0}
\put(0.2,2.85){1}
\put(0.2,3.15){0}
\put(0.2,5.45){1}
\put(0.2,5.75){0}
\put(0.2,8.05){1}
\end{picture}
\end{center}
\caption{Comparison between spatial spectra $Q_{u}\left(x,k_{z}\right)$ computed from the experiments ($-$) at station $x_{\cal {C}}=0.05$~m and the theoretical expression ($-$) computed from (\ref{eq:Qdiscrete_Aconstant}) on the boundary $x=0$, for $W=\lambda_{e}$ (top),  $W=2\lambda_{e}$ (middle) and $W=3\lambda_{e}$ (bottom).}
\label{fig:spectrawidth}
\end{figure}

The normalized experimental spatial Fourier spectra $Q_{u}\left(x_{\cal{C}},k_{z}\right)$ at $x_{\cal{C}}=0.05$~m and the normalized theoretical spectra $Q_{u}\left(0,k_{z}\right)$ are presented in figure~\ref{fig:spectrawidth}. A standout feature of the results is that the theoretical Fourier transform does a remarkably good job of predicting the spectrum of the experimental wave field. For $W/\lambda_{e}=1$, the spectrum is broadly centered around the expected vertical wavenumber $k_{e}=79.9$~m$^{-1}$, and this principal peak becomes increasingly sharp for $W/\lambda_e=2$ and 3. This evolution is quantified by the half width $\delta\/k$, defined as the width of the principal spectral peak at half peak amplitude, the values of which are listed in table \ref{table:updown_emission} for the three different configurations.

Another notable feature of the spectra is that strength of the upward propagating wavefield ($k_z<0$) significantly decreases with increasing $W/\lambda_{e}$.
Computing the parameter $\beta_{d}$ for experimental and numerical cases 1, 7 and 8 of table~\ref{table:exp} quantifies this trend. As seen in table~\ref{table:updown_emission}, 98\% of the energy propagates in the desired direction for $W/\lambda_e\geq\/2$. Furthermore, we computed $\delta\/k/k_e$ and $\beta_{d}$ for cases 4, 9 and 10 of table~\ref{table:exp} too, and the results presented in table \ref{table:updown_emission} show that the influence of the width is more significant than the influence of the angle of emission.

\begin{table}
\centering
\begin{tabular}{c|ccc|ccc|ccc|ccc}
 & \multicolumn{3}{|c|}{$\delta k/k_{e}$ (15$^\circ$)} & \multicolumn{3}{|c|}{$\beta_{d}$ (15$^\circ$)}& \multicolumn{3}{|c|}{$\delta k/k_{e}$ (45$^\circ$)} & \multicolumn{3}{|c}{$\beta_{d}$ (45$^\circ$)} \\
$W/\lambda_e$ & num. & exp. & theo. &  num. & exp. & theo. & num. & exp. & theo. & num. & exp. & theo. \\[3pt]
$1$ & $1.67$ & $1.11$ & $1.21$ & $0.97$ & $0.94$ & $0.94$ & 1.31 & 1.08 & 1.21 & 0.97 & 0.92 & 0.94 \\
$2$ & $0.58$ & $0.57$ & $0.60$ & $0.99$ & $0.97$ & $0.96$ & 0.61 & 0.58 & 0.60 & 0.99 & 0.96 & 0.96 \\
$3$ & $0.37$ & $0.39$ & $0.40$ & $0.99$ & $0.99$ & $0.97$ & 0.40 & 0.39 & 0.40 & 0.99 & 0.98 & 0.97 \\
\end{tabular}
\caption{Relative half width, $\delta k/k_{e}$, and relative energy of the downward propagating wave, $\beta_d$, for cases 1,  7 and 8 of table \ref{table:exp}, corresponding to a propagation angle of 15$^\circ$, and experiments 4, 9 and 10 for a propagation angle of 45$^\circ$.}
\label{table:updown_emission}
\end{table}

\subsubsection{Discretization}
\label{subsct:discretized}

To study the impact of spatially discretized, rather than continuous, forcing, experiments were performed for $W/\lambda_e=3$ with $M=4$ and $M=12$; these being experiments 12 and 1 in table \ref{table:exp}. For comparison, corresponding numerical simulations were also performed for these two configurations, as well as for the idealized case $M \rightarrow \infty$, which is listed as experiment 11 in table \ref{table:exp} and corresponds to forcing discretized on the scale of the grid resolution in the numerical simulations.

\begin{figure}
\begin{center}
\begin{picture}(13.0,6.5)
\put(0.1,0.0){\includegraphics[width=13cm]{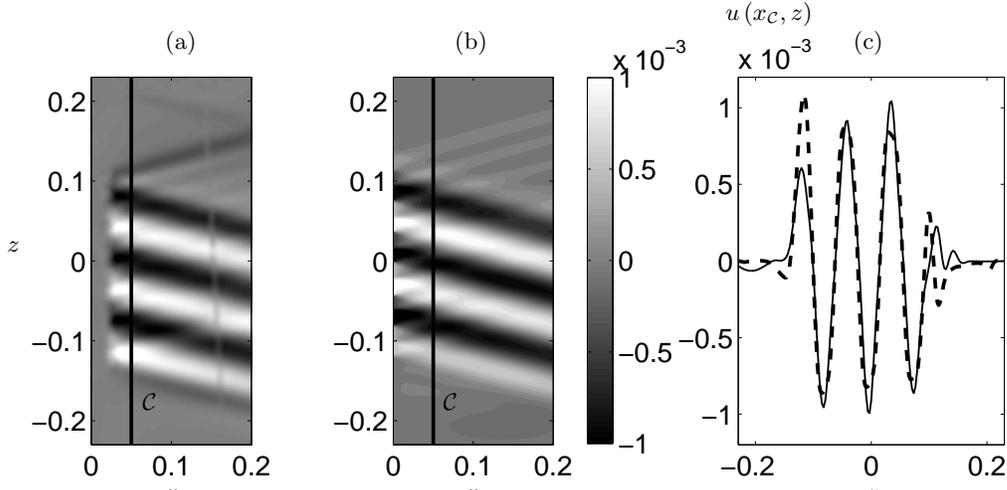}}
\put(9.35,6.08){$u\left(x_{\cal{C}},z\right)$}
\put(1.9,5.7){(a)}
\put(5.75,5.7){(b)}
\put(11.05,5.7){(c)}
\put(1.9,-0.3){$x$}
\put(5.9,-0.3){$x$}
\put(11.225,-0.3){$z$}
\put(-.2,3){$z$}
\put(1.6,0.9){${\cal {C}}$}
\put(5.6,0.9){${\cal {C}}$}
\end{picture}
\end{center}
\caption{Horizontal velocity fields for (a) experiment and (b) simulation for case 12 in Table~\ref{table:exp} with $M=4$. (c) Vertical profiles along the cut $\cal{C}$ located at $x_{\cal {C}}=0.05$~m for the experiment ($--$) and the simulation ($-$).  All lengths are in m and all velocities in m~s$^{-1}$.}
\label{fig:discrete_planewaves}
\end{figure}

Snapshots of the experimental and numerical wave fields for $M=4$ are presented in figures~\ref{fig:discrete_planewaves}(a) and \ref{fig:discrete_planewaves}(b), respectively, while figure~\ref{fig:discrete_planewaves}(c) presents vertical cuts of the horizontal velocity field at $x_{\cal {C}}=0.05$~m for these two data sets. Even for this coarse discretization, there is still a remarkably smooth and periodic wave field that looks little different to that obtained using $M$=12 (see figure \ref{fig:finiteextent}). And once again there is good agreement between experiment and numerics, with the slight exception of the outer edges of the profile where the numerical peaks are a little larger amplitude.

\begin{figure}
\begin{center}
\begin{picture}(10.0,9.2)
\put(0,0.25){\includegraphics[width=10cm]{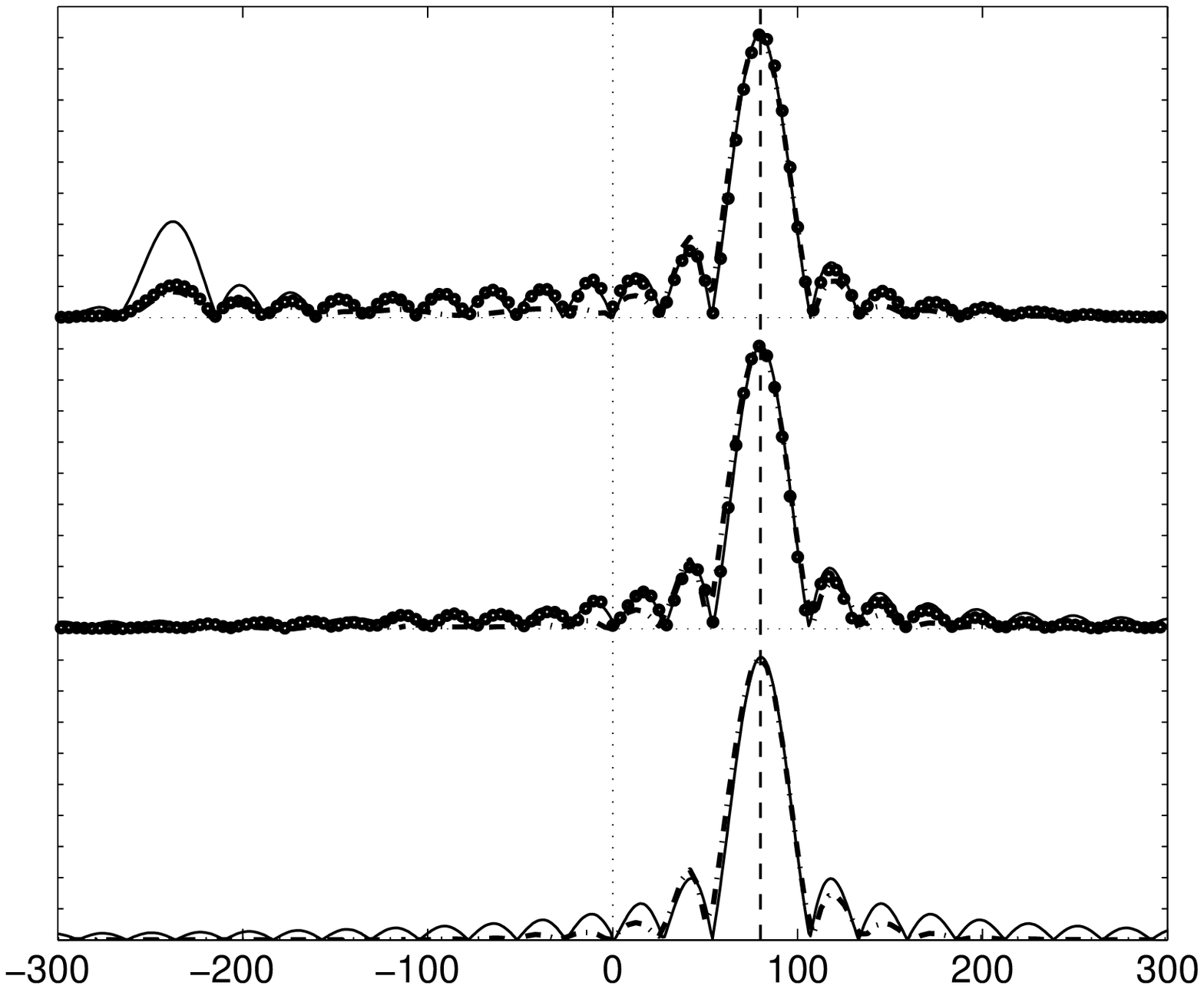}}
\put(5,-0.25){$k_z$}
\put(-.2,8.75){$Q_u(x,k_z)/\max_{k_z}Q_u(x,k_z)$}
\put(6.2,-0.25){$k_{e}$}
\put(6.31,0){\vector(0,1){0.58}}
\put(0.2,0.55){0}
\put(0.2,2.85){1}
\put(0.2,3.15){0}
\put(0.2,5.45){1}
\put(0.2,5.75){0}
\put(0.2,8.05){1}
\end{picture}
\end{center}
\caption{Comparison between spatial spectra $Q_{u}\left(x,k_{z}\right)$ computed from the experiments ($\circ$) and numerical simulations ($\mathbf{-\cdot}$) at station $x_{\cal {C}}=0.05$~m ,and the theoretical expression ($-$) computed from (\ref{eq:Qdiscrete_Aconstant}) on the boundary $x=0$, for $M=4$ (top), 12 (middle) and $\infty$ (bottom).}
\label{fig:discrete_planewavesspectra}
\end{figure}

Although the cross-section of the emitted downward-propagating, nominally plane wave looks reasonable, the discretization does induce more of an undesired upward-propagating wave,  which can clearly be seen in figures~\ref{fig:discrete_planewaves}(a) and~\ref{fig:discrete_planewaves}(b). Fourier spectra for experiments, numerics and theory corresponding to  $M$=4, 12 and $\infty$ are presented in figure~\ref{fig:discrete_planewavesspectra} and, as predicted by~(\ref{eq:Qdiscrete_Aconstant}), the strength of the negative wave numbers noticeably increases with decreasing $M$. Most notably, for $M=4$ (corresponding to $\ell=19.6$~mm) the discretization introduces a peak around $k_z=-235(\pm 4)$~m$^{-1}$, which is strongest in the theoretical spectrum but nevertheless evident in the experimental and numerical spectra. By analogy with the theory of optical gratings, this value is in good agreement with the canonical formula $2\pi/\lambda_e - 2\pi/\ell = -241$~m$^{-1}$, which can also be inferred from~(\ref{eq:Qdiscrete_Aconstant}) when $\ell\ll\lambda_e$. For all three cases the principal peak remains sharp, with $\delta\/k/k_e =0.42 $. The value of $\beta_d$ is 0.96 when $M=4$, so a vast majority of the energy is still in the downward propagating wave field.

\subsubsection{Amplitude}
\label{subsct:amplitude}

All the results presented thus far have been for $A_{0}=5.0$~mm. To investigate the impact of a significantly larger amplitude of forcing on the quality of the radiated wave field, we performed a series of experiments with the same parameters as experiments~1, 3 and 4, with the exception of $A_{0}=35.0$~mm; they are listed as experiments 13 to 15 in table \ref{table:exp}.

We found that the qualitative level of agreement between experiment and numerics for snapshots of the wave field was comparable to that presented in figure~\ref{fig:finiteextent}(a) and (b). When a more quantitative comparison is made, however, some consequences of the higher-amplitude forcing become apparent.
For example, for vertical cuts located at \mbox{$x_{\cal {C}}=0.075$~m} the amplitude of the horizontal velocity component in the experiments was $4.4~\pm~0.9$~mm~s$^{-1}$, compared to $5.50~\pm~0.25$~mm~s$^{-1}$ in the numerical simulations. We also note that although the forcing amplitude was increased by a factor of~7, in the experiments the wave amplitudes only increased by a factor of around 5. \footnote{We had to make these, and later, comparisons, for a vertical cross section further away from the generator than in our previous studies because the much larger amplitude motion of the plates created a more intense wave field very close to the generator, where it was not possible to get reliable experimental data.}

\begin{figure}
\begin{center}
\begin{picture}(13,7.7)
\put(0,0.25){\includegraphics[width=13truecm]{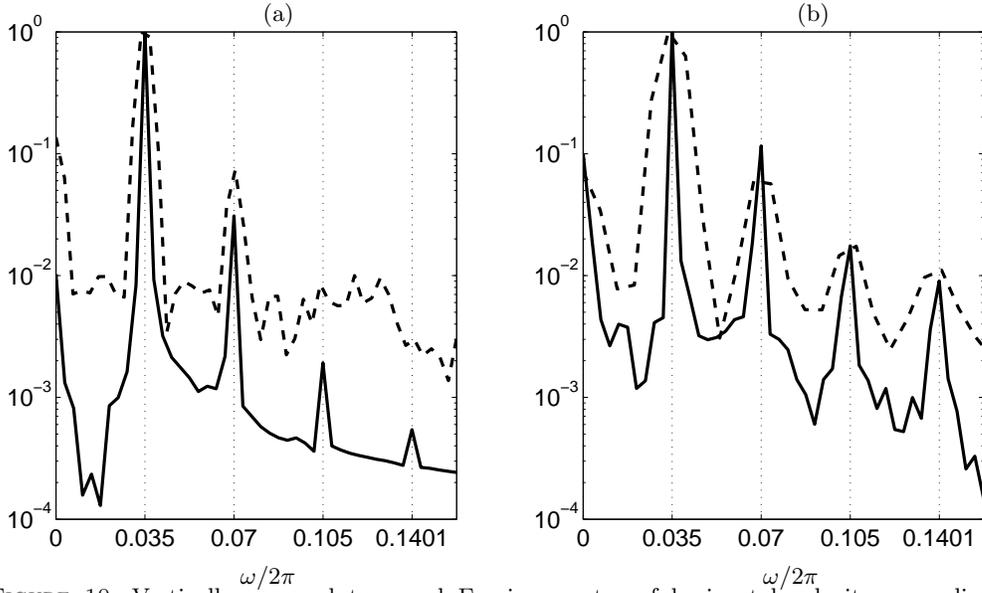}}
\put(3.1,-0.25){$\omega/2\pi$}
\put(10.05,-0.25){$\omega/2\pi$}
\put(10.5,7.25){(b)}
\put(3.4,7.25){(a)}
\end{picture}
\caption{Vertically-averaged temporal Fourier spectra of horizontal
velocity normalized by their maximum amplitude. The data was obtained from cuts at $x_{\cal {C}}=0.075$~m for (a) $A_{0}=0.005$~m and (b) $A_{0}=0.035$~m, for experiments ($--$) and numerics ($-$).}\label{fig:amplitude_Tspectra}
\end{center}
\end{figure}

Figure~\ref{fig:amplitude_Tspectra} presents vertically-averaged temporal spectra of the horizontal velocity for the cuts at $x_{\cal {C}}=0.075$~m, for both small and large amplitude forcing. The numerical data, being at somewhat higher temporal resolution, has a lower noise level than the experimental data. Both experimental and numerical spectra display the same qualitative change; the large amplitude forcing introduces more significant higher harmonic content into the wave field.

\begin{figure}
\begin{center}
\begin{picture}(10,9.2)
\put(0,0.25){\includegraphics[width=10cm]{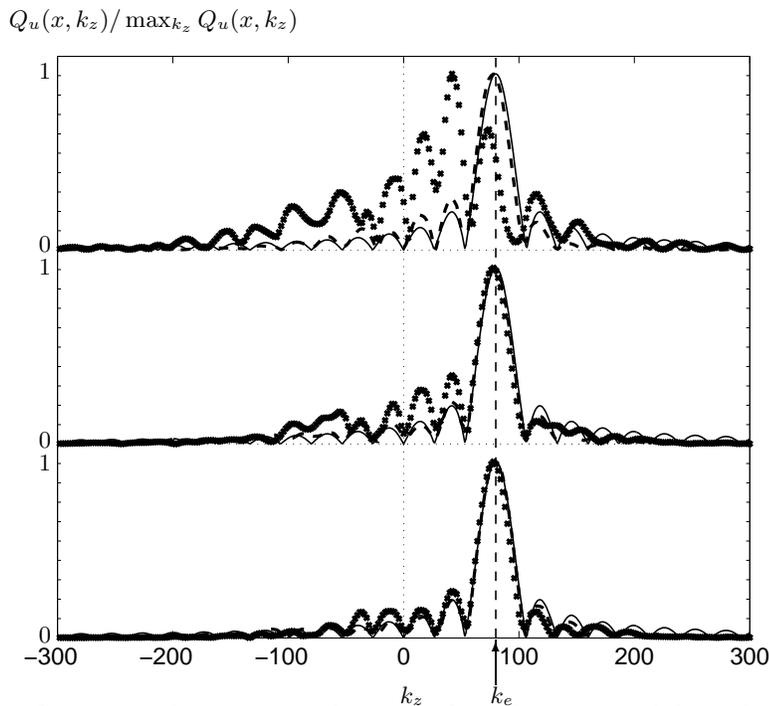}}
\put(5,-0.25){$k_z$}
\put(-.2,8.75){$Q_u(x,k_z)/\max_{k_z}Q_u(x,k_z)$}
\put(6.2,-0.25){$k_{e}$}
\put(6.27,0){\vector(0,1){0.58}}
\put(0.2,0.55){0}
\put(0.2,2.85){1}
\put(0.2,3.15){0}
\put(0.2,5.45){1}
\put(0.2,5.75){0}
\put(0.2,8.05){1}
\end{picture}
\caption{Comparison between spatial spectra $Q_{u}\left(x,k_{z}\right)$ computed from the experiments with $A_{0}=0.035$~m ($\times$) and $A_{0}=0.005$~m ($\mathbf{--}$) at station $x_{\cal {C}}=0.075$~m and the theoretical expression ($-$) computed from (\ref{eq:Qdiscrete_Aconstant}) on the boundary $x=0$, for $\theta=45^{\circ}$ (top), $30^{\circ}$ (middle) and $15^{\circ}$ (bottom).}\label{fig:amplitude_Zspectra}
\end{center}
\end{figure}

The normalized spatial spectra of the wave field for frequencies corresponding to $\theta=15$, $30$ and $45^{\circ}$ are presented for both small and large amplitude forcing in figure~\ref{fig:amplitude_Zspectra}. The angle of emission does not seem to significantly impact the quality of the wave field for the small amplitude forcing, but this is not so for the larger amplitude forcing, for which we find that $\beta_d$ decreases from $0.97$ to $0.94$ for $\theta=15^{\circ}$ and $30^{\circ}$, respectively (although $\delta k/k_{e}=0.38$ remains constant). For $\theta=45^{\circ}$ and $A_{0}=35.0$~mm, the generator no longer generates a clean plane wave, the main peak being centered around $41.5$~m$^{-1}$ and $\beta_d=0.80$. The cause of this breakdown is not easy to discern. One hypothesis is that breakdown occurs at criticality, when the angle of wave propagation exceeds the maximum slope angle of the face of the generator, $\theta_{m}=\frac{\pi}{2}-\arctan\left(\frac{2\pi A_{0}}{\lambda_{e}}\right)$. The reason is not as simple as this, however, since $\theta_m=20^{\circ}$ for the experiments with $A_{0}=35.0$~mm, and yet the generator is still an efficient source of plane waves for $\theta=30^{\circ}$. Other factors, such as the nonlinear coupling between the plates and the wave field, as characterized by the Reynolds number of the plate motion, would also seem to play a role.

\subsection{Summary}

Through a systematic series of experiments, listed in Table~\ref{table:exp}, we can draw several conclusions about the ability of the novel wave generator to generate plane waves. We find that the spatial Fourier transform of the profile of the wave generator can reasonably predict \emph{a priori} the spectrum of the radiated wave field. If more comprehensive resources are available, a numerical simulation with boundary forcing applied at $x$=0 can reliably reproduce the emitted wave field for small-amplitude forcing. As one might expect, the spectrum of the wave field becomes increasingly sharp about the dominant wavelength, and thus more akin to a plane wave, as the number of wavelengths excited increases, and even a very crude spatial discretization of the desired wave form  produces a remarkably smooth and coherent wave field. For large amplitude forcing, the main impact is an increase of the harmonic content of the wave field.  Overall, we conclude that a vertically-standing wave generator produces a radiated wave field of high quality provided $\theta\leq\/45^{\circ}$.

\section{Wave Beams}
\label{sct:stevenson}

Wave beams are a common feature of internal wave fields in both laboratory experiments \cite[][]{bib:PeacockTabei05,bib:GostiauxDauxois07} and geophysical settings \cite*[][]{bib:Lametal04,bib:Martinetal06},  since they are readily generated by periodic flow relative to an obstacle, be it a cylinder or an ocean ridge for example. We choose to investigate the so-called Thomas--Stevenson profile \cite[][]{bib:ThomasStevenson72}, a viscous self-similar solution of (\ref{eq:master_equations}) that can be considered as the far-field limit of the viscous solution of elliptic cylinder oscillating in a stratified fluid \cite[][]{bib:HurleyKeady97}. It has been shown that this profile describes oceanographically relevant internal wave beams far from their generation site at the continental shelf \cite[][]{bib:GostiauxDauxois07}; after their reflection at the bottom of the ocean, such wave beams are thought to be
the cause of solitons generated at the thermocline \cite[see][for instance]{bib:NewDaSilva02, bib:Gerkema01}.

\subsection{Analysis}

Consider a downward-propagating, right-going beam at angle $\theta\in[0,\pi/2]$ with respect to the horizontal, here $\theta$ being defined to be positive in a clockwise sense. Let $\xi=x\cos\theta-z\sin\theta+l$ and $\eta=x\sin\theta+z\cos\theta$ be the coordinates parallel and transverse to the wave beam, respectively, with $l$ corresponding to the distance from the point source to the origin of the cartesian frame at the center of the active region of the generator. At leading order, the parallel and transverse velocity components and buoyancy fields of the Thomas--Stevenson profile are:
\begin{subequations}
\begin{equation}
u^{\prime}\left(\xi,\eta,t\right)= u_0 \left(\frac{\xi N^2\sin\theta}{g}\right)^{-2/3}\!\!\!\!\!\!\Re\left\{\int_{0}^{\infty}k\exp\left(-k^{3}\right)\exp\left(\mathi k\alpha \frac{\eta}{\xi^{1/3}}-\mathi\omega t\right)\,\mbox{d}k\right\}, \label{eq:TS_v_longitudinal}
\end{equation}
\begin{equation}
v^{\prime}\left(\xi,\eta,t\right)= u_0 \left(\frac{\xi N^2\sin\theta}{g}\xi\alpha^{3/2}\right)^{-2/3}\!\!\!\!\!\!\!\!\!\Re\left\{-\mathi \int_{0}^{\infty}\!\!\!\!\!k^{3}\exp\left(-k^{3}\right)\exp\left(\mathi k\alpha \frac{\eta}{\xi^{1/3}}-\mathi\omega t\right)\!\!\mbox{d}k\right\}, \label{eq:TS_v_transverse}
\end{equation}
and
\begin{equation}
b\left(\xi,\eta,t\right)= N u_0\left(\frac{\xi N^2\sin\theta}{g}\right)^{-2/3}\!\!\!\!\!\!\Re\left\{-\mathi \int_{0}^{\infty}k\exp\left(-k^{3}\right)\exp\left(\mathi k\alpha \frac{\eta}{\xi^{1/3}}-\mathi\omega t\right)\,\mbox{d}k\right\}, \label{eq:TS_buoy}
\end{equation}
\label{eq:thomas_stevenson}
\end{subequations}
where $u_0$ is the amplitude of the horizontal velocity and $\alpha=\left(2N\cos\theta/\nu\right)^{1/3}$.

In principle, to most accurately reproduce (\ref{eq:thomas_stevenson}) one should tilt the generator and configure the profile of the forcing plates to match the transverse profile of the Thomas-Stevenson beam. As stated earlier, however, we consider a vertically-standing wave generator, since one will typically want to investigate several wave beam angles in an experiment and reconfiguring the system for each angle is impractical.
Thus we seek to reproduce the profile (\ref{eq:thomas_stevenson}) by using only the longitudinal velocity profile (\ref{eq:TS_v_longitudinal}), instead of the true horizontal velocity profile, to prescribe the forcing at $x$=0, i.e.
\begin{equation}
u\left(0,z,t\right)=u^{\prime}\left(l,z,t\right).
\label{eq:TS_forcing}
\end{equation}
This is an approximation of the exact solution, which will become increasingly valid with decreasing $\theta$. For a given viscosity, stratification and forcing frequency that determine $\alpha$, equation~(\ref{eq:TS_forcing}) sets effective values for the parameters $\ell$ and $u_0$ in (\ref{eq:thomas_stevenson}).

\subsection{Configuration}
\label{subsct:TSsetups}

The experiments were performed using the ENS Lyon generator (see table~\ref{table:generator}) with a background  stratification $N=0.82$~rad~s$^{-1}$ and forcing frequencies $\omega=0.20$~rad~s$^{-1}$, 0.44~rad~s$^{-1}$ and 0.58~rad~s$^{-1}$, corresponding to propagation angles of 14, 32 and 44$^\circ$ respectively. The arrangement used twenty-one plates to discretize the profile at $x=0$, thirteen of which covered the $9$~cm active region. Based on the results of section~\ref{subsct:discretized}, this level of discretization is expected to be
sufficient to resolve the structure of the wave beam. The configuration of the cams (amplitude and evolution of the phase) is
depicted in figure \ref{fig:excentricityandcames}, with the maximum amplitude of oscillation being $10$~mm.

Experimental visualizations were performed using the Synthetic Schlieren method, which gives direct measurements of the gradient of the buoyancy field. For the following study, we integrate this data and compare the measured buoyancy field with the analytical model (\ref{eq:TS_buoy}) and numerical results. For these simulations, the numerical domain was $0.80$~m long and $0.60$ to $1.01$~m high, with a vertical resolution $\Delta z=0.79$~mm. The forcing on the boundary was discretized on the scale of the numerical grid.

\subsection{Results}
\label{subsct:TSresults}

\begin{figure}
\begin{center}
\begin{picture}(13,4.8)
\put(0.0,0.0){\includegraphics[width=13truecm]{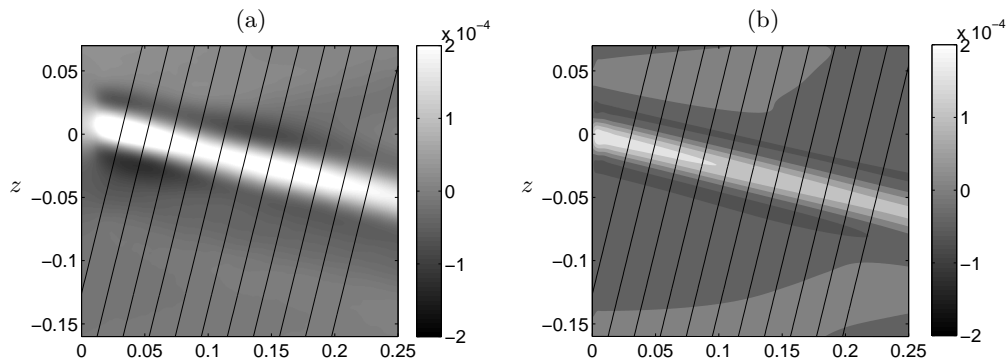}}
\put(2.7,4.4){(a)}
\put(9.5,4.4){(b)}
\put(-0.3,2.2){$z$}
\put(6.5,2.2){$z$}
\put(2.7,-0.3){$x$}
\put(9.5,-0.3){$x$}
\end{picture}
\end{center}
\caption{Buoyancy field $b(x,z,t)$ for a $14^\circ$ Thomas--Stevenson beam. Black lines are the transverse cuts used in figure~\ref{fig:TS_transverse_numexp} to compare with the model. (a) Experimental and (b) numerical buoyancy fields are in rad.s$^{-2}$.}
\label{fig:TS_rho_cuts}
\end{figure}

A direct comparison between experimental and numerical buoyancy fields for $\theta=14^\circ$ is presented in figure~\ref{fig:TS_rho_cuts}. There is close qualitative agreement between the two, and it is notable that there is no visible sign of any upward propagating beam coming from the generator, due to the highly resolved, smooth boundary conditions that were used. For a vertical cut at $x_{\cal{C}}=0.05$~m of the vertical component of the density gradient, $\partial_zb$, we find that $\beta_d=0.98$ for the experiments and $\beta_d=0.99$
for the numerics. We note that the maximum values of $\partial_z b$ are around 2\% of the background stratification,
and so the wave field can reasonably be considered linear. Analysis of temporal spectra confirmed that harmonics were at
least one order of magnitude smaller than the fundemantal signal.

\begin{figure}
\begin{center}
\begin{picture}(13.5,4.3)
\put(1,0.1){\includegraphics[width=5.5truecm]{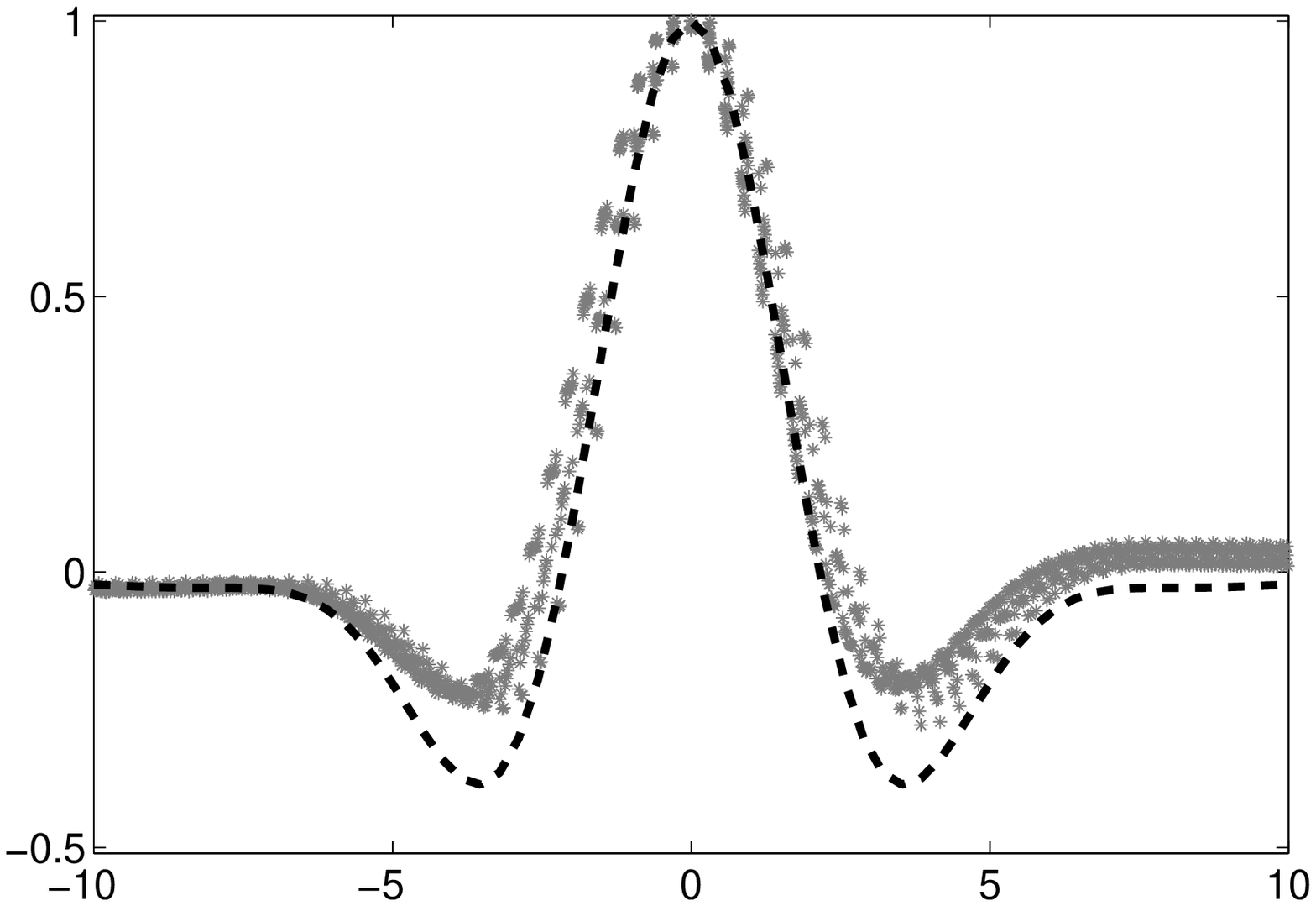}}
\put(7.8,0.1){\includegraphics[width=5.5truecm]{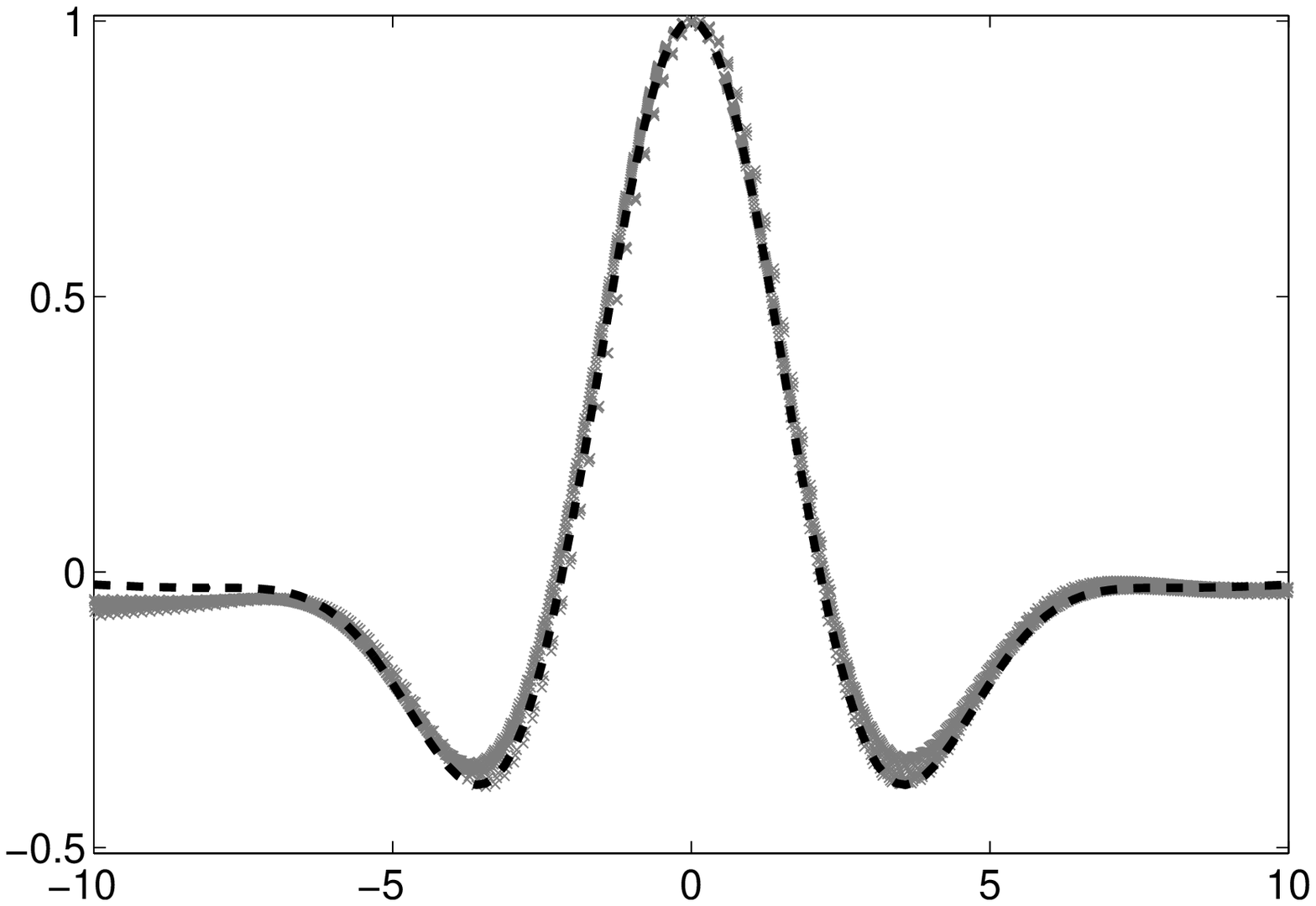}}
\put(3.65,4.){$(a)$}
\put(10.45,4.){$(b)$}
\put(-0.15,1.9){$\Re\left({b}/{b_{m}}\right)$}
\put(6.7,1.9){$\Re\left({b}/{b_{m}}\right)$}
\put(3.55,-0.3){$\alpha\eta/\xi^{1/3}$}
\put(10.35,-0.3){$\alpha\eta/\xi^{1/3}$}
\end{picture}
\end{center}
\caption{Real parts of the normalized transverse profiles $b/b_{m}$ as defined in (\ref{eq:TS_transverse2}) for (a) experiments and (b) numerical simulations as a function of $\alpha\eta/\xi^{1/3}$ for the case $\theta=14^\circ$. Asterisks correspond to the different cuts in figure~\ref{fig:TS_rho_cuts}, the dashed line corresponds to the analytical solution in (\ref{eq:TS_transverse2}).}
\label{fig:TS_transverse_numexp}
\end{figure}

For a more quantitative comparison with the self-similar solution of Thomas--Stevenson, we investigate the buoyancy field extracted along the cuts indicated in figures~\ref{fig:TS_rho_cuts}(a) and (b). Specifically,
we consider the normalized transverse profiles:
\begin{equation}
\frac{b({\eta}/{\xi^{1/3}})}{b_{m}(\xi)} = \frac{\displaystyle\int_{0}^{\infty}k\exp\left(-k^{3}+ik\alpha{\eta}/{\xi^{1/3}}\right)\,\mbox{d}k}{\displaystyle \int_{0}^{\infty}k\exp\left(-k^{3}\right)\,\mbox{d}k}\,, \label{eq:TS_transverse2}
\end{equation}
where
\begin{equation}
b_{m}(\xi)= \max_{\eta}\,b\left(\xi,\eta,0\right)= N u_0\left(\frac{\xi N^2\sin\theta}{g}\right)^{-2/3}\!\!\!\!\!\!\,\,\,\int_{0}^{\infty}k\exp\left(-k^{3}\right)\,\mbox{d}k \label{eq:TS_b_m2}
\end{equation}
is the maximum amplitude of the buoyancy perturbation (\ref{eq:TS_buoy}) along a transverse cut, which lies at the center of the beam ($\eta=0$). The results are presented in figure~\ref{fig:TS_transverse_numexp}(a) and (b), where it can be seen that both the experimental and numerical results evolved spatially in a self-similar manner, with only small differences compared to the analytical model.

We also confirmed the ability of the generator to produce waves beams for some steeper angles, by performing experiments for $\theta=32^\circ$ and $\theta=44^\circ$. Comparisons with the analytical solution revealed the same level of agreement as for the $\theta=14^\circ$ wave beam. The wave generator still emitted a beam in only one direction, and the experimental values of $\beta_d$ were $0.99(8)$ for $\theta=32^\circ$ and $0.99(8)$ for $\theta=44^\circ$, with no discernable change in the value of $\beta_d$ for the numerical simulations.

\subsection{Summary}

On the basis of this study, we conclude that the novel wave generator is capable of producing a wave beam structure of a desired form, which demands excitation of a prescribed Fourier spectrum. 
We have demonstrated this for the example of the Thomas--Stevenson profile, and speculate that although it is not necessary, perhaps even closer agreement with theory
can be obtained using a tilted generator.

\section{Vertical modes} \label{sct:mode}

Vertical internal wave modes play an important role in our current understanding of internal tides in the ocean \cite[][]{bib:GarrettKunze07}.
To date, however, there has been little progress in producing high-quality vertical modes in laboratory experiments. \cite{bib:Thorpe68} generated a mode-1 disturbance by oscillating a flap hinged about a horizontal axis at mid-depth; and \cite{bib:Nicolaou93} produced low modes in an essentially two-layer system with a thermocline, but neither of these approaches can readily produce arbitrary modes in an arbitrary stratification.
Generalized forcing of a spectrum of vertical modes was obtained by \cite{bib:Echeverrietal09}, who used an oscillating Gaussian topography to generate an internal wave field, and developed a robust algorithm for extracting modal amplitudes from experimental data. Here, we demonstrate the ability of the novel wave generator to reliably produce arbitrary internal wave modes.

\subsection{Analysis}

For a stratified fluid of constant $N$, the horizontal velocity field associated with the $n^{\mathrm{th}}$ vertical mode of frequency $\omega$ propagating from left to right is:
\begin{equation}
u_{n}(x,z,t) = \mathrm{Re}\left[u_{0} \cos\left(\frac{n\upi
z}{H}\right)\exp\left(\mathrm{i}\frac{n\upi}{H\cot\theta}\,
x-\mathrm{i}\omega t\right)\right] \,, \label{eq:mode1forcing}
\end{equation}
where $u_0$ is a complex amplitude that sets both magnitude and
phase, $n$ is a positive integer, $z=0$ and $z=H$ are the bottom and
top boundaries, respectively, and $\theta$ is the first-quadrant
angle that satisfies the dispersion relation (\ref{eq:dispersion_relation}). Note that $\theta$ does not specify
the direction of energy propagation for a vertical mode and only plays a part in setting the horizontal wavenumber \cite*[][]{bib:Gill}.

The idealized boundary forcing of the horizontal velocity at $x=0$ to
excite the $n^{\mathrm{th}}$ vertical mode is:
\begin{equation}
u(0,z,t) = a\cos\left(\frac{n\upi z}{H}\right)\cos (\omega t)\,,
\label{eq:mode_boundary}
\end{equation}
where $a$ is an arbitrary amplitude. To analyze the quality of the wave field generated by this forcing, instead of Fourier transforms one uses modal analysis, which is equivalent to Fourier series in a constant stratification. The horizontal velocity component of the resulting wave field can be written as:
\begin{equation}
u(x,z,t) = \sum_{n=1}^{n=\infty} a_{n} \cos\left(\frac{n\upi
z}{H}\right)\cos\left(\frac{n\upi x}{H\cot\theta} - \omega t +
\phi_{n}\right) \,, \label{eq:modal_decomp}
\end{equation}
where $a_{n}$ and $\phi_{n}$ are the strength and the phase of the
$n^{\mathrm{th}}$ mode respectively.
Similar results exist for other physical variables, including vertical velocity and buoyancy fields.

From a practical point of view, one decomposes the experimental
generated wave field at a fixed $x$ location into the vertical basis
modes using the numerical algorithm described and implemented in
\cite{bib:Echeverrietal09}. The modal decomposition is then
performed at several other $x$ locations, and the variations in
$a_{n}$ and $\phi_{n}$ across various $x$ locations gives an
estimate of the experimental errors in the results. One can
reliably correct for viscous dissipation of the modes, if needs be, by introducing in (\ref{eq:modal_decomp}) a multiplicative term of the type $e^{-f_n x}$, with
\begin{equation}
f_n=\frac{\nu n^3}{2\omega}\left(\frac{\pi}{H} \right)^2 \left[ \frac{N^2}{N^2-\omega^2}\right]^2 \label{eq:viscous_mode}
\end{equation}
being the spatial damping rate \cite[][]{bib:Thorpe68,bib:Echeverrietal09}. In our experiments, $f_n$ is very small (roughly $10^{-4}$~m$^{-1}$ for mode-1 and $10^{-3}$~m$^{-1}$ for mode-2) and hence viscous dissipation can be neglected.

The discretization results for plane waves in section \ref{subsct:discretized} suggest that there could be limitations on the ability to resolve a vertical mode due to the discretization of the wave generator. By computing the modal decomposition of the discrete plate arrangement, we find that provided at least $3n$ equispaced plates are used to represent the $n^{\mathrm{th}}$ mode, then more than $95\%$ of the boundary forcing is contained in the $n^{th}$ mode.
To ensure good quality of the generated wave field also requires one to account for the approximation that the boundary forcing of horizontal velocity occurs at a fixed $x$ location even though the plates are actually moving.
For this approximation to hold, the ratio of the maximum amplitude of oscillations of the plates to the horizontal wavelength corresponding to the $n^{\mathrm{th}}$ mode should be much smaller than unity, i.e. ${nA_{0}}/({2H\cot\theta}) \ll 1$. Provided these two conditions are satisfied, a high-quality wave field is to be expected.

\subsection{Configuration}
The MIT facility was used for the experiments with $N_{p}=64$ of the total $82$ plates (see table~\ref{table:generator}).
Individual vertical modes were produced by configuring the $N_p$ plates of the wave generator to reproduce (\ref{eq:mode_boundary}). The amplitude of oscillation of the $j^{th}$ plate centered at vertical position $z_{j}=j\ell/2$ of mode $n$ was $A(z_{j})=A_0\cos(n\upi z_j/H)$, with $\ell$ being the plate thickness (see Fig.~\ref{fig:excentricityandcames} for more details of the configuration). Experiments were performed for modes~1 and 2.
The spatial resolution of the forcing, relative to the vertical wavelength of the mode being forced, was $1/64$ in both experiments.
The fluid depth was $H=0.416$~m, the maximum amplitude of oscillation was $A_0=2.5$~mm and the stratification was
$N\simeq0.85$~rad~s$^{-1}$. Visualization of the wave field was performed using PIV technique. Using this arrangement it was possible
to observe the wave field in a~45~cm-wide horizontal domain and covering the full depth of the tank. No corresponding numerical simulations were performed in this case.

\subsection{Results}
\label{modes_results}

\begin{figure}
\begin{center}
\begin{picture}(12.0,5.8)
\put(-0.1,0.){\includegraphics[width=0.9\linewidth]{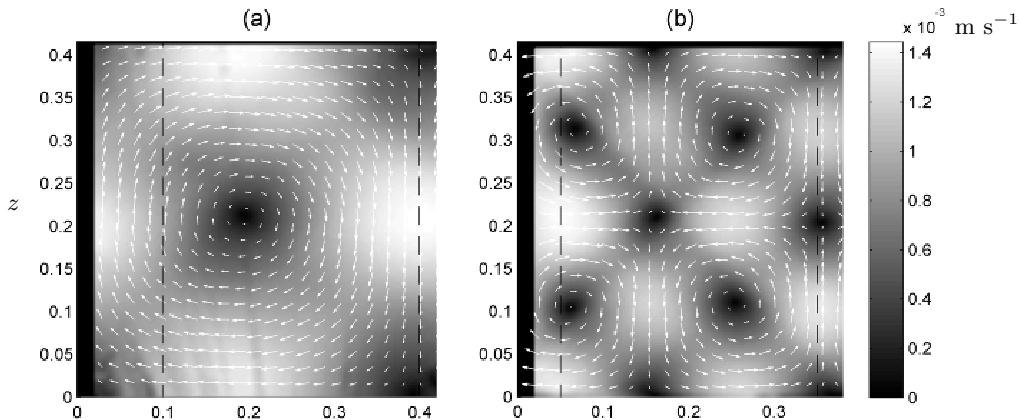}}
\put(-0.5,2.85){$z$}
\put(2.7,-0.3){$x$}
\put(8.45,-0.3){$x$}
\put(12.1,5.2){m s$^{-1}$}
\end{picture}
\caption{Snapshots of experimental velocity fields for (a) mode-1 and (b) mode-2. The location of the wave generator is $x=0$ and
the vertical dashed lines bound the domain over which modal decomposition is performed. The $x$ and $z$ coordinates are in meters and
the grey scale is the velocity magnitude in m~s$^{-1}$. Arrows indicate local velocity direction.}
\label{fig:mode1_45degrees}
\end{center}
\end{figure}

We present detailed results for mode-1 and mode-2 wave fields with $\theta$=45$^\circ$,  for which the horizontal
wavelengths excited were $k_x=7.55$~m$^{-1}$ and $k_x=15.1$~m$^{-1}$, respectively. Snapshots of the
wave fields obtained in the vicinity of the generator are presented in figure
\ref{fig:mode1_45degrees}. One can clearly see the characteristic
structure of a single vortex that covers the entire vertical domain for mode-1 (figure~\ref{fig:mode1_45degrees}(a)), whereas for mode-2 the structure comprises
stacked pairs of counter rotating vortices (figure~\ref{fig:mode1_45degrees}(b)).

\begin{figure}
\begin{center}
\begin{picture}(12.0,5.0)
\put(0.0,0.0){\includegraphics[width=0.9\linewidth]{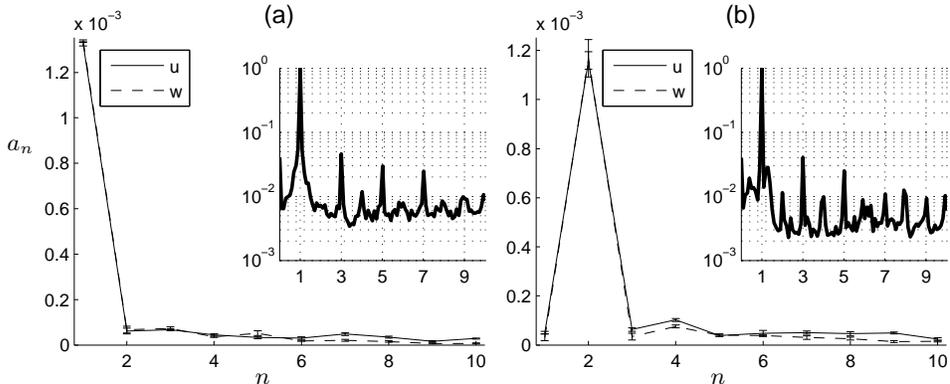}}
\put(-0.5,2.9){$a_n$}
\put(2.8,-0.2){$n$}
\put(8.85,-0.2){$n$}
\end{picture}
\caption{Modal decompositions of the wave fields presented in (a) figure \ref{fig:mode1_45degrees}(a), and (b) figure \ref{fig:mode1_45degrees}(b).
Error bars are the standard deviation of modal amplitude for the $81$ vertical cross-sections studied between the dashed lines in figure \ref{fig:mode1_45degrees}. The insets
show vertically averaged absolute value of the temporal Fourier spectra of $u$ at $x=10$~cm and $x=5$~cm in
(a) and (b), respectively.}
\label{fig:mode2_45degrees}
\end{center}
\end{figure}

Modal decomposition of the experimental wave fields was performed at $81$ $x$-locations in the regions bounded
by the vertical dashed lines in figures \ref{fig:mode1_45degrees}(a) and \ref{fig:mode1_45degrees}(b), and the results are presented in figure~\ref{fig:mode2_45degrees}. Since the maximum
values of both $u$ and $w$ were the same for these $\theta=45^\circ$ wave fields, reliable values of $a_{n}$ and $\phi_{n}$
were obtained from both components of the velocity field. The two wave fields were clearly dominated by their mode number, with by far the largest detectable
amplitude being for modes-1 and modes-2 in the two respective experiments, and with very little variability across the experimental domain, emphasizing the
weak impact of viscosity in these experiments. We also observed very little variability in the phase of the dominant mode across the visualization window, implying that the wave
fields were highly spatially coherent. Since energy flux scales as $a_n^2/n$ (\cite{bib:Echeverrietal09}) over $98.8$~\% of the energy was in the desired mode in each experiment.
The efficiency of conversion, defined as $a_{n}/A\omega$, was $0.89$ and $0.78$ for the two experiments.
The insets in figure \ref{fig:mode2_45degrees} present the vertically-averaged temporal Fourier spectrum of~$u$. These are dominated by the fundamental frequency,
demonstrating that very little higher-harmonic content was generated by nonlinearity.

\subsection{Summary}

Overall, these experiments reveal that the novel wave generator is capable of producing very high-quality radiating vertical modes in a linear stratification. For smaller angles (i.e. lower
frequencies), $\theta=15^{\circ}$ for example, we found that the wave field took longer
horizontal distances to evolve into established modal solutions, presumably due to the longer horizontal wave lengths of the modes.
For higher frequencies, $\theta=60^{\circ}$ for example, we observed high-quality modal solutions within our visualization window, but
with a small variation in the dominant modal strength for mode-2, possibly due to nonlinear effects. For higher angles, therefore, it would seem prudent to decrease the maximum
amplitude of oscillations and/or reduce the stratification $N$ (allowing for lower forcing frequency) in an effort to produce as clean a wave field as possible. In principle, this approach can be extended to experiments with nonlinear density stratifications, provided the stratification is known {\em a priori} so that the generator can be configured for the appropriate vertical structure of the horizontal velocity field.

\section{Conclusions}
\label{sct:conclusion}

Through combined experimental, numerical and theoretical studies we have demonstrated that the novel type of internal wave
generator, comprising a series of stacked, offset plates, can reliably shape the spatial structure of an experimental
internal wave field and enforce wave propagation. This approach is similar in spirit to multiple-paddle techniques that have been developed for generating {\em surface}
waves \cite[see][for instance]{bib:Naito06}. We have demonstrated the ability of the
generator to produce three qualitatively-different types of wave field: plane waves, wave beams and vertical modes. This new technology therefore provides a very useful tool to study all manner of internal wave scenarios in the laboratory, in order to
gain insight into geophysically important problems. Furthermore, our studies reveal that the Fourier transform of the spatial
profile of the wave generator provides a reasonably accurate prediction of the form of emitted wave field, making it a useful
tool when designing experiments.

There are numerous examples of where this new found capability can now be utilized. For example, observations in several
locations, in particular in the Bay of Biscay \cite[][]{bib:NewPingree90,bib:NewDaSilva02}, have determined that an
internal tidal beam striking the thermocline is responsible for the generation of solitons. A full understanding of the
generation mechanism has yet to be achieved, however, and laboratory experiments using the Thomas--Stevenson beam profile
impinging on a thermocline could provide significant insight. Indeed, the interaction of wave beams with nonlinear features in the
density stratification is of widespread interest \cite[][]{bib:MathurPeacock09}, since this is also relevant to how and where atmospheric
internal waves break and deposit their momentum \cite[][]{bib:Nault07}. In regards to ocean mixing problems, an open issue
is to determine the fate of the internal tide which, among other things, can be scattered by topography~\cite[][]{bib:Johnston1,bib:Ray}.
The ability to directly generate vertical modes provides a new capability to study these important processes in controlled settings.
Other interesting avenues for research are the generation of shear waves (McEwan \& Baines 1974) and extensions to three-dimensional wave fields, which could be achieved by introducing some horizontal spatial structure (in the $y$-direction) to the leading edge of the moving plates.

\section*{Acknowledgments}

We thank D. Le Tourneau, M. Moulin and A. Gallant for technical help, and Robert Smith and Ian Curtis for their creative input and decisive inspiration. This work has been partially supported by the ANR grant PIWO (ANR-08-BLAN-0113-01), the MIT-France program, NSF grant 0645529 and ONR grant N00014-09-0282.

\bibliographystyle{jfm}
\bibliography{OS}

\end{document}